\documentclass[aps,prd,twocolumn,showpacs,nofootinbib,superscriptaddress,amsmath,amssymb,showkeys,showpacs]{revtex4-1}
\usepackage{graphicx}
\usepackage{dcolumn}
\usepackage{bm}

\newcommand{\be}{\begin{equation}}
\newcommand{\ee}{\end{equation}}
\newcommand{\bqa}{\begin{eqnarray}}
\newcommand{\eqa}{\end{eqnarray}}
\begin{document}
\title{Strong interaction between kaons in the reactions $e^+ e^- \to K^+ K^- \gamma$ and $e^+ e^- \to K^0 \bar{K^0} \gamma$}
\author{L.~Le\'sniak}
\email[Electronic address:]{leonard.lesniak@ifj.edu.pl}
\affiliation{Institute of Physics, Jagiellonian University, 30-348 Krak\'ow, Poland}
\author{F. Sobczuk}
\affiliation{Institute of Physics, Jagiellonian University, 30-348 Krak\'ow, Poland}
\author{M. Silarski}
\affiliation{Institute of Physics, Jagiellonian University, 30-348 Krak\'ow, Poland}
\author{F.~Morawski}
\affiliation{Institute of Physics, Jagiellonian University, 30-348 Krak\'ow, Poland}

\begin{abstract}
A theoretical model of the reactions $e^+ e^- \to K^+ K^- \gamma$
and $e^+ e^- \to K^0 \bar{K^0} \gamma$ has been derived. 
The strong interaction between kaons is
taken into account using a general form of the $K \bar{K}$ scattering amplitude.
It is shown that some models formulated in the past are particular cases of the present approach.  
The formulae for the $K \bar{K}$ effective mass dependence of the differential cross section as well as for the angular kaon and photon
distributions and for the branching fractions of the $\phi(1020) \to K^+ K^- \gamma$ and $\phi(1020) \to K^0 \bar{K^0}\gamma$ decays have been obtained.
We present numerical results for the functions entering into   
transition amplitudes, $K \bar{K}$ effective mass distributions, total cross sections, and branching fractions. 
Finally, the model is generalized to treat other reactions with two
pseudoscalar mesons accompanying a photon in the final state. 
\end{abstract}
\maketitle
\section{Introduction}
\label{Introduction}
In the standard classification of the quark-antiquark states the scalar meson nonet is not yet uniquely 
determined and the scalar resonances constitute the least known group of mesons (for reviews see, for example, Refs.~\cite{pdg2017,Klempt}).  
There are many different hypotheses about their internal structure.
Besides the interpretation as $q\bar{q}$ mesons~\cite{Morgan},
these particles were also proposed to be the $qq\bar{q}\bar{q}$ tetraquark states~\cite{Jaffe},
mixtures of $q\bar{q}$ and meson-meson systems~\cite{Beveren} or even quarkless gluonic
hadrons ~\cite{Johnson}.
Since both $f_{0}$(980) and $a_{0}$(980) masses are very close to the sum of the $K$ and ${\bar K}$
masses, they have been considered as $K\bar{K}$ quasi-bound states~\cite{Weinstein}.
Verification of these hypotheses requires a precise information about the elastic and inelastic
amplitudes of the kaon-antikaon interaction.
Existing data from the $K \bar K$ phase analyses are not yet sufficiently precise even to make a statement whether the kaon-antikaon interaction in the $S$-wave near the threshold is attractive or repulsive~\cite{cohen,etkin}.

Since the kaonic targets are not available and the colliding kaon beams do not exist as well, the only way
to study the kaon-kaon interaction is a production of kaon pairs
with low relative momenta and analysis of their rescattering. This type of studies was conducted, 
for example, in the $pp \to ppK^+K^-$  reaction at the COSY synchrotron in J\"ulich,
Germany. However, small cross sections and the presence of strongly interacting protons in the final state,
made it impossible to accurately estimate the scattering length of the $K^+K^-$
interaction~\cite{Ye,Silarski:2013rfa,dzyuba}.
Thus, it seems that it is much better to study less complicated final states produced, for example, in heavy meson decays $J/\psi \to \phi K^+K^-$~\cite{bes},
$B^0_s \to J/\psi f_0(980)$~\cite{Aaij:2011fx} or in the  $e^+e^- \to K^+K^-\gamma$ and $e^+ e^- \to K^0 \bar{K^0} \gamma$ reactions.  
The latter processes are suitable to study the strong interactions between kaon pairs since no other hadrons exist in the final state. 

Radiative $\phi(1020)$ decays into $K \bar K \gamma$ states via the intermediate production of scalar mesons $f_0(980)$ and $a_0(980)$ have been theoretically studied already in the late eighties and in nineties of the 20th century~\cite{Nussi,Ivan,Lucio1,Bramon,Bramon1,Close,Lucio2,KL,Oller1,Marco}.
Then, after a construction of the $\Phi$-factory in Italy at Frascati
other papers appeared~\cite{Bramon2,Markushin,Aczasow2001,Oller2,Kala,Escribano,Black,Gokalp1,Gokalp2,Eidelman}. 
The KLOE Collaboration has searched for the decay $\phi(1020) \to K^0 {\bar K}^0 \gamma$
and obtained an upper limit of the branching fraction equal to $1.9 \cdot 10^{-8}$~\cite{kloe2}.
For a full description of the kaon-antikaon interaction also other coupled decay channels
have to be considered.
Thus KLOE has carried out a series of analyses~\cite{kloe2008} to study the properties of scalar mesons
in the $\pi^0\pi^0\gamma$, $\eta\pi^0\gamma$ 
and
$\pi^+\pi^-\gamma$ final states~\cite{kloe0,kloe1,KLOE}.
Earlier, the $\phi$ radiative decays into $\pi^+\pi^-\gamma$,    $\pi^0\pi^0\gamma$ and $\eta\pi^0\gamma$ have been measured in Novosibirsk by SND and CMD-2 Collaborations (see, for example, Refs.~\cite{Ak1,Ak2,Ac1,Ac2}).
 However, there are no data for the $\phi(1020)$ decay into $K^+K^-\gamma$, even the branching ratio is not known. This decay could be studied, in principle, also using the KLOE data set.

Three theoretical models have been used in the analysis of the KLOE data for the $e^+e^- \to \pi^+ \pi^- \gamma$ reaction~\cite{KLOE}. In the first model 
a phenomenological parametrization of the reaction amplitude by a suitable combination of the 
elastic pion-pion amplitude and the transition amplitude from two kaons in the intermediate state
to two pions in the final state is made~\cite{mod1}. A direct $\phi$ meson coupling to
the $f_0(980)$ resonance and a photon without any assumption about the internal structure of the
scalar meson is postulated in the second model~\cite{NS}. The total and differential cross sections
for the  $e^+e^- \to K^+K^-\gamma$ process predicted by this model can be found in~\cite{konf}~\footnote{The parameters of this model, called the no-structure model, have been taken from Table 1 of Ref.~\cite{KLOE} (label NS).}.
An essential ingredient of the third model, called the kaon-loop model, is a formation of a loop of two charged
kaons to which photons can couple~\cite{mod3}. The best fits to data~\cite{KLOE} have been obtained for the former
two models. It turned out, however, that important quantities like the $f_0(980)$ mass
and the coupling constants of that resonance to the $K^+K^-$ and  $\pi^+\pi^-$ pairs, obtained 
from fits to data, are in mutual disagreement.

One of the reasons for this unsatisfactory situation
could lie in lack of information on the $K^+K^- \gamma$ channel. Moreover, one should try to describe
simultaneously all the mentioned coupled channels.
Transition amplitudes of the reactions coupled to the $K^+K^-$ channel were described, for example, in Refs.~\cite{kam1,KLL,kam3,Furman}.
As it will be shown later, they can be incorporated in a more general scheme taking into account all the relevant channels.
Some threshold properties of the $K \bar K$ amplitudes have been already examined in~\cite{KL95} and
a new parametrization of the resonant production amplitudes near inelastic thresholds has been proposed in Refs.~\cite{LL} and~\cite{L2009}. 
These results could be useful in analyses of the experimental data
in which the $K \bar K$ pairs are present the final state.

  In the construction of the theoretical models of the reaction $e^+ e^- \to K^+ K^- \gamma$ one should take into account an important role playing by scalar mesons $f_0(980)$ or $a_0(980)$ in an intermediate production step:
$ e^+e^- \to \phi(1020) \to \gamma (f_0 + a_0) \to K^+ K^- \gamma$. 
At the  $e^+ e^-$ center-of-mass energies close to 1 GeV the $\phi(1020)$ meson contribution to the reaction cross-section is dominant. 
The $\phi$ meson decays most frequently into a pair of charged kaons, the corresponding branching fraction is 49.2 \%~\cite{pdg2017}. 
The charged kaons can interact strongly and electromagnetically emitting photons. 
Thus the kaon-loop mechanism has been frequently used to construct models of the reaction amplitude.
 
Very often in literature the scalar mesons have been treated as point-like particles (see, for example, Refs.~\cite{Nussi,Ivan,Lucio1,Bramon,Lucio2}).
However, Close, Isgur and Kumano in Ref.~\cite{Close} have modeled the scalar mesons as extended objects, formed of the $K\bar K$ quasi bound molecules. 
It has been found that the branching fraction for the decay $\phi(1020) \to f_0 \gamma$ depends sensitively on the molecule radius which in turn is related to the $K\bar K$ binding energy. 
The model presented in Ref.~\cite{Close} is essentially nonrelativistic in treatment of the $K \bar K$ interaction but the reaction amplitude for the radiative $\phi$ decay can be "relativized" in some manner explained in the article.

In Ref.~\cite{KL}, using a specific distribution of the relative $K \bar K$ momenta given in Ref.~\cite{Close},
numerical calculations of the branching fractions and the mass spectra for the radiative $\phi$ decays into scalar mesons $a_0$ and $f_0$ with subsequent
decays of $a_0$ into $\pi \eta$ and $K^+K^-$ and $f_0$ into $\pi \pi$ and $K^+K^-$ have been performed. 
The relevant $K \bar K$ molecule radius was equal to 1.2 fm (see Eq. (4.13) in Ref.~\cite{Close}).

In 2001 Achasov and Gubin calculated the differential cross sections for the 
reactions $e^+ e^- \to \phi \to K^+ K^- \gamma$ and $e^+ e^- \to \phi \to K^0 \bar K^0 \gamma$ ~\cite{Aczasow2001}. In this case the $K^+K^-$ loop function from Ref.~\cite{Ivan} has been used.
Formally, this choice corresponds to the limit of vanishing $K \bar K$ molecule radius.

In the above mentioned models of the reaction amplitude the $K\bar K$ scattering amplitude has always a resonant character due to a presence of scalar mesons in the intermediate states. 
We know, however, that the $K\bar K$ amplitude can have an additional non-resonant part which is not included in the approaches discussed above.
So, the idea is to extend the existing models of the radiative $\phi$ decays into $K\bar K \gamma$ by inclusion of a more general form of the $K\bar K$ scattering amplitude.
Our aim is to construct a model which will be suitable in the coupled-channel
analyses of different radiative $\phi$ meson decays.

As the first step the amplitude for the $e^+ e^- \to K \bar K \gamma$ reaction with the charged kaon loop is considered. 
Other contributions to the reaction amplitudes like the loops of other particles than the $K^+K^-$ ones can be added later in future model developments.

The paper is structured as follows.
In Sec.~\ref{model} 
we formulate an extended version of the theoretical model describing the $e^+ e^- \to K^+K^- \gamma$ reaction. Then in Subsection~\ref{limitqzero} we discuss the limit of vanishing photon energy. In Subsection~\ref{otherappr} we show examples of models which are particular cases of the model described in Sec.~\ref{model}. 
In Sec.~\ref{KKamplitudes} some properties of the on-shell and off-shell
elastic $K^+ K^-$ amplitudes are outlined. 
 The formulae for the differential cross sections and for the branching fraction of the decay $\phi(1020) \to K^+K^- \gamma$ are written in Sec.~\ref{Diff}.
In Sec.~\ref{numK+K-} we give the results of numerical calculations performed for this reaction. 
In Sec.~\ref{other} we briefly show how to generalize the model to the description of other reactions with two pseudoscalar meson pairs accompanying photon in the final-state.
Sec.~\ref{DK0K0} is devoted to an application of the model to the 
$e^+ e^- \to K^0 {\bar K}^0 \gamma$ channel.
 The conclusions are given in Sec.~\ref{Concl}.
Finally, in Appendix A an amplitude approximation is explained and in Appendix B   
some longer formulae are gathered.

\section{ 
Theoretical model of the amplitude for the reaction $e^+ e^- \to K^+ K^- \gamma $}
\label{model}
\subsection{Model derivation}
 \label{model derivation}
Below we present a derivation of the theoretical model for the reaction $e^+ e^- \to K^+ K^- \gamma$. 


One can start from a set of three amplitudes describing the so-called final state radiation (FSR) process.  
In the lowest order of quantum electrodynamics a virtual photon is initially emitted from the incoming $e^+e^-$ pair (see Fig.~1 a), then the final photon can be coupled to the vertex connecting $K^+$, $K^-$ and that virtual photon, or directly from $K^+$ or $K^-$ (Fig.~1 b,c). 
The diagram in Fig.~1(a) represents the so-called contact term which is needed to satisfy the gauge  invariance of the FSR amplitude.
The form of the FSR amplitude is well known. For example, for the 
$e^+ e^- \to \pi^+ \pi^- \gamma$ reaction the corresponding expression for the FSR amplitude can be found in Eq.~(1) of Ref.~\cite{Czyz}.

\begin{figure}[h]
\centering
\includegraphics[width=4.2cm]{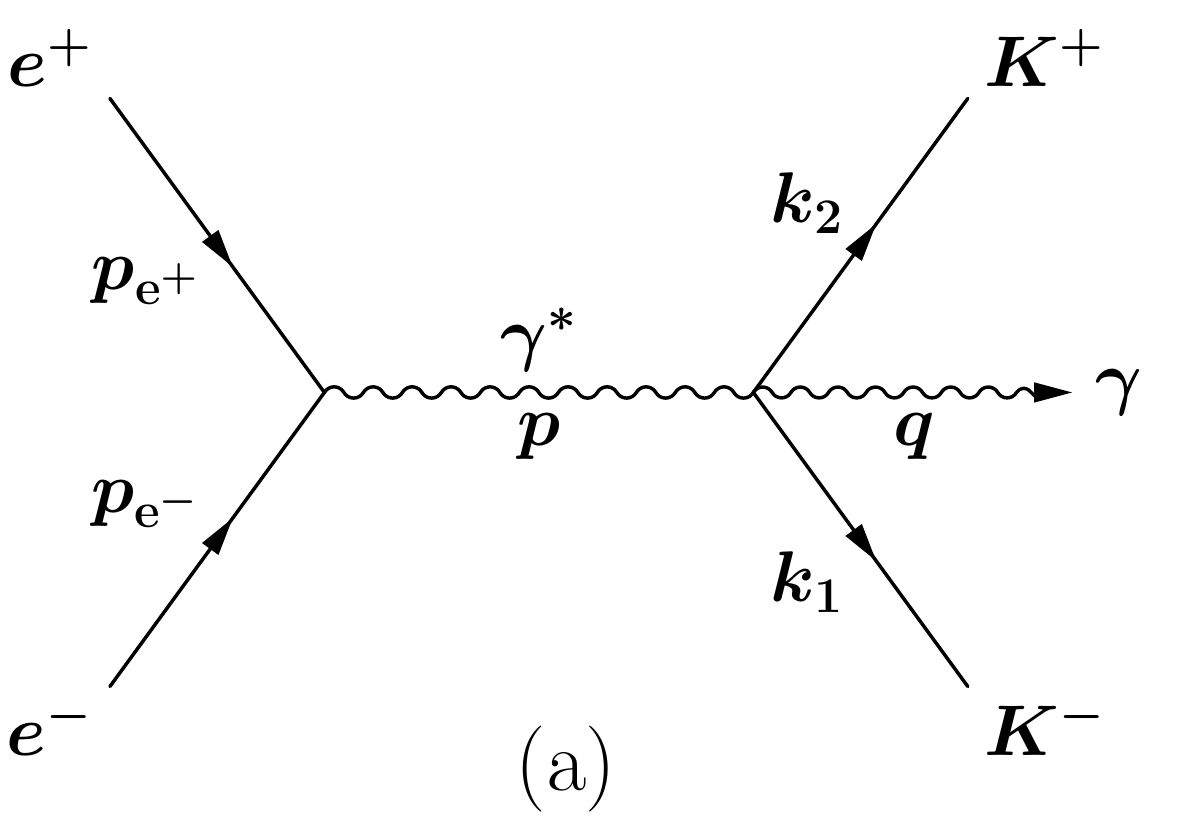}
\includegraphics[width=4.2cm]{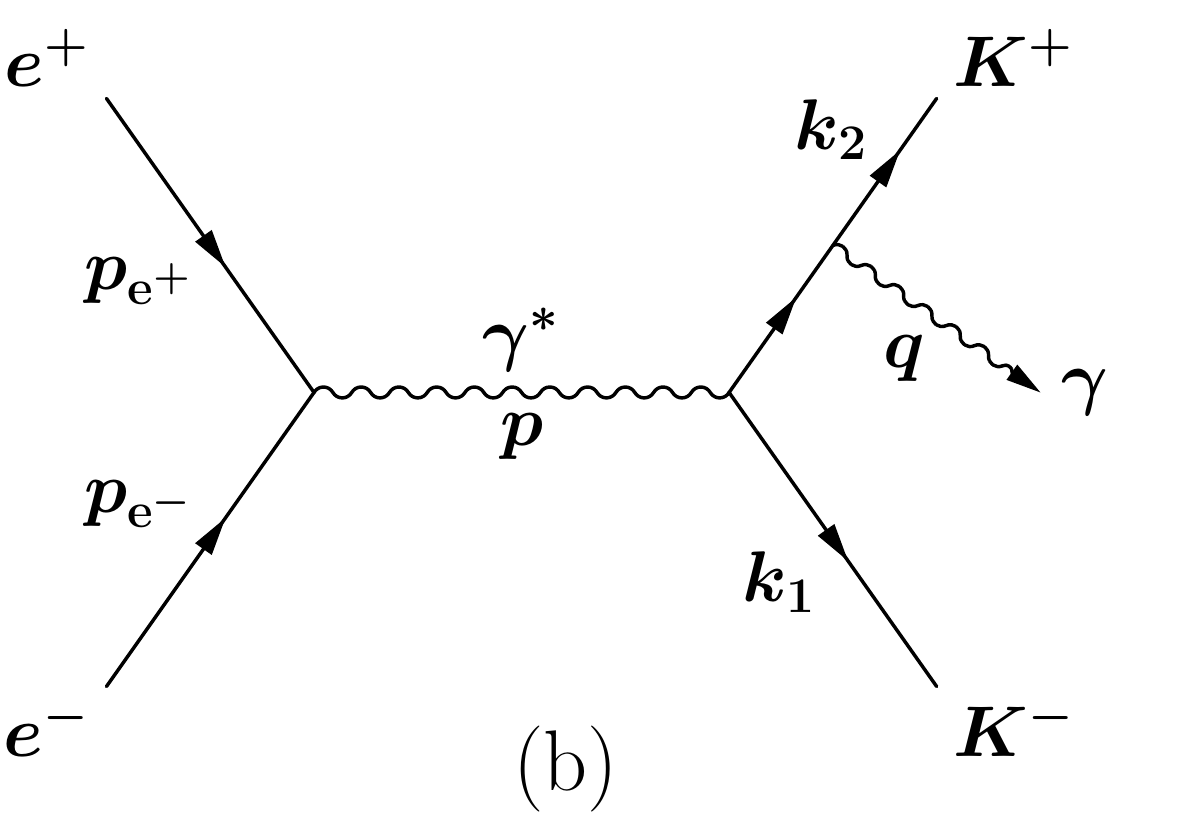}
\includegraphics[width=4.2cm]{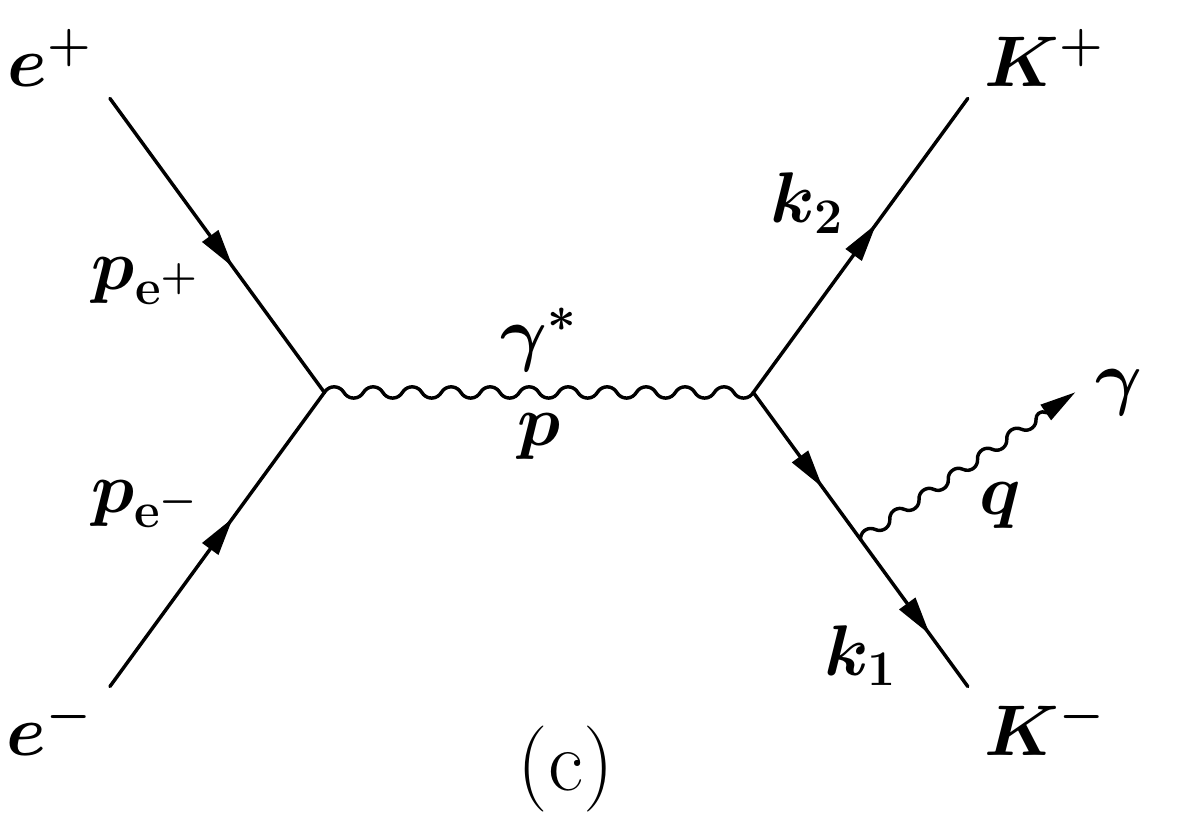}
\caption{Diagrams corresponding to the final-state radiation in the $e^+ e^- \rightarrow K^+ K^- \gamma$
reaction.
\label{fig-1}}       
\end{figure}
 
\begin{figure}[ht]
\centering
\includegraphics[width=4.2cm]{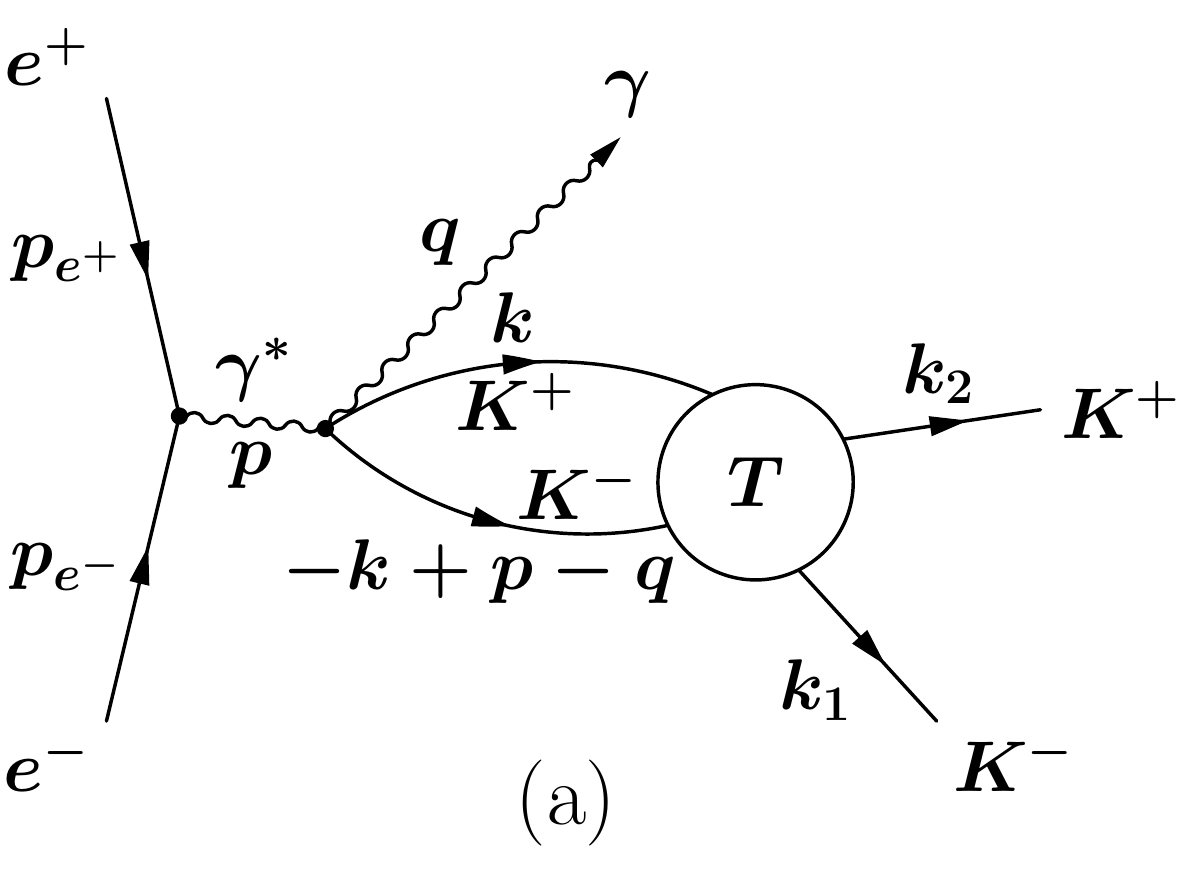}~~
\includegraphics[width=4.2cm]{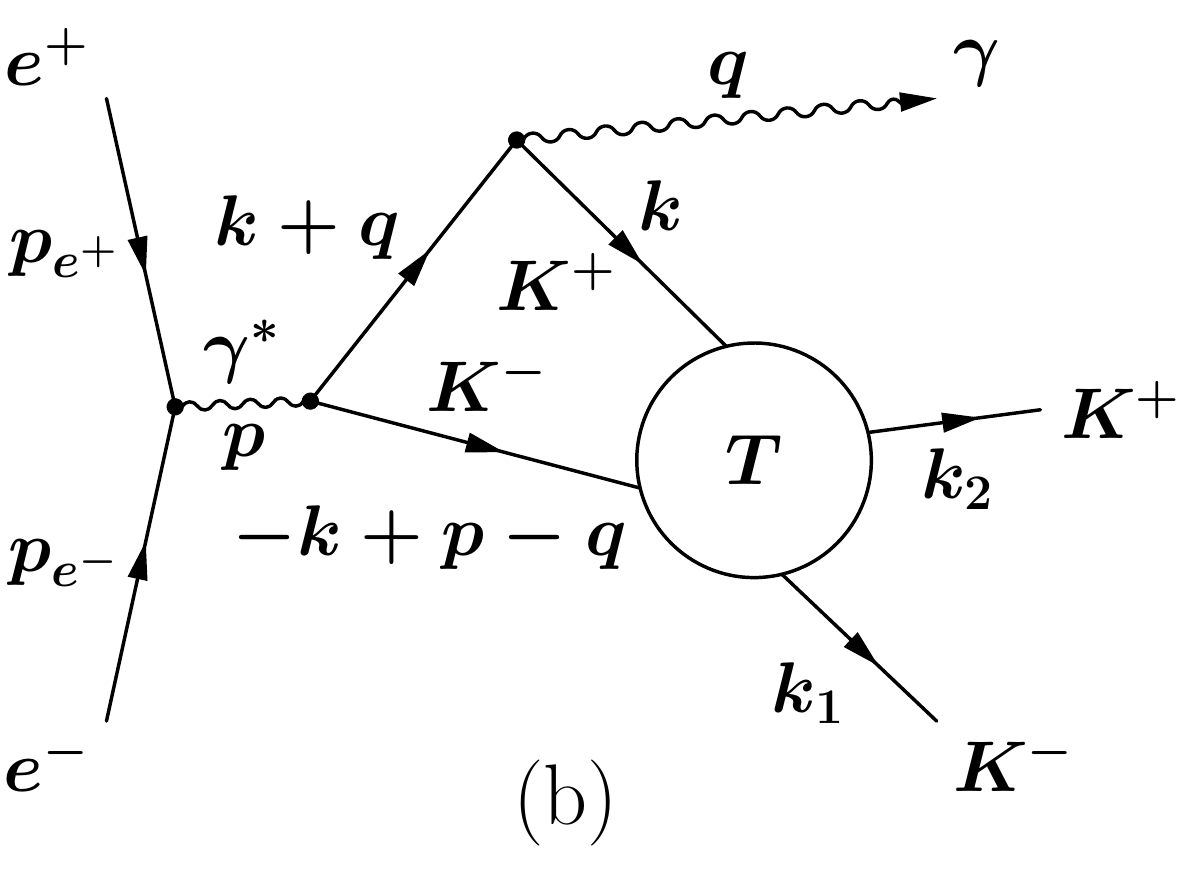}\\
\includegraphics[width=4.2cm]{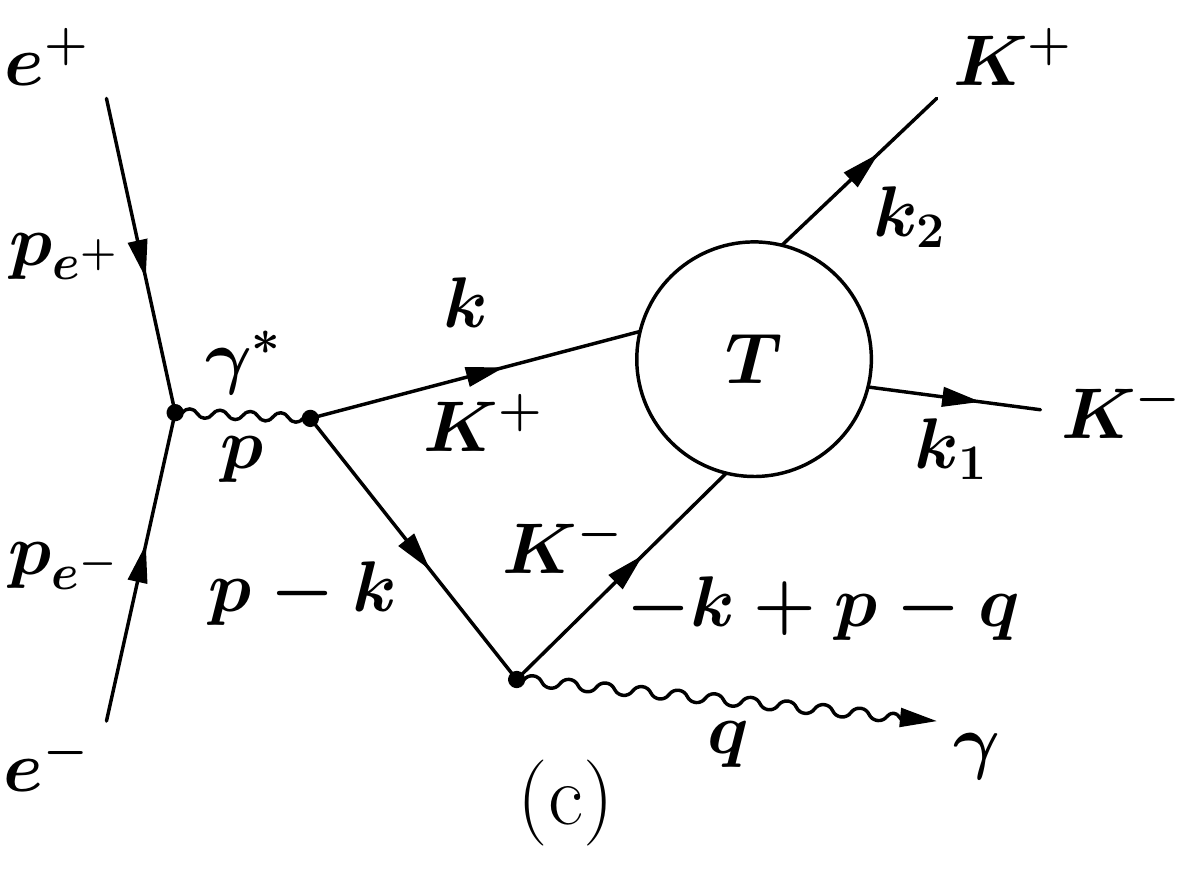}~~
\includegraphics[width=4.2cm]{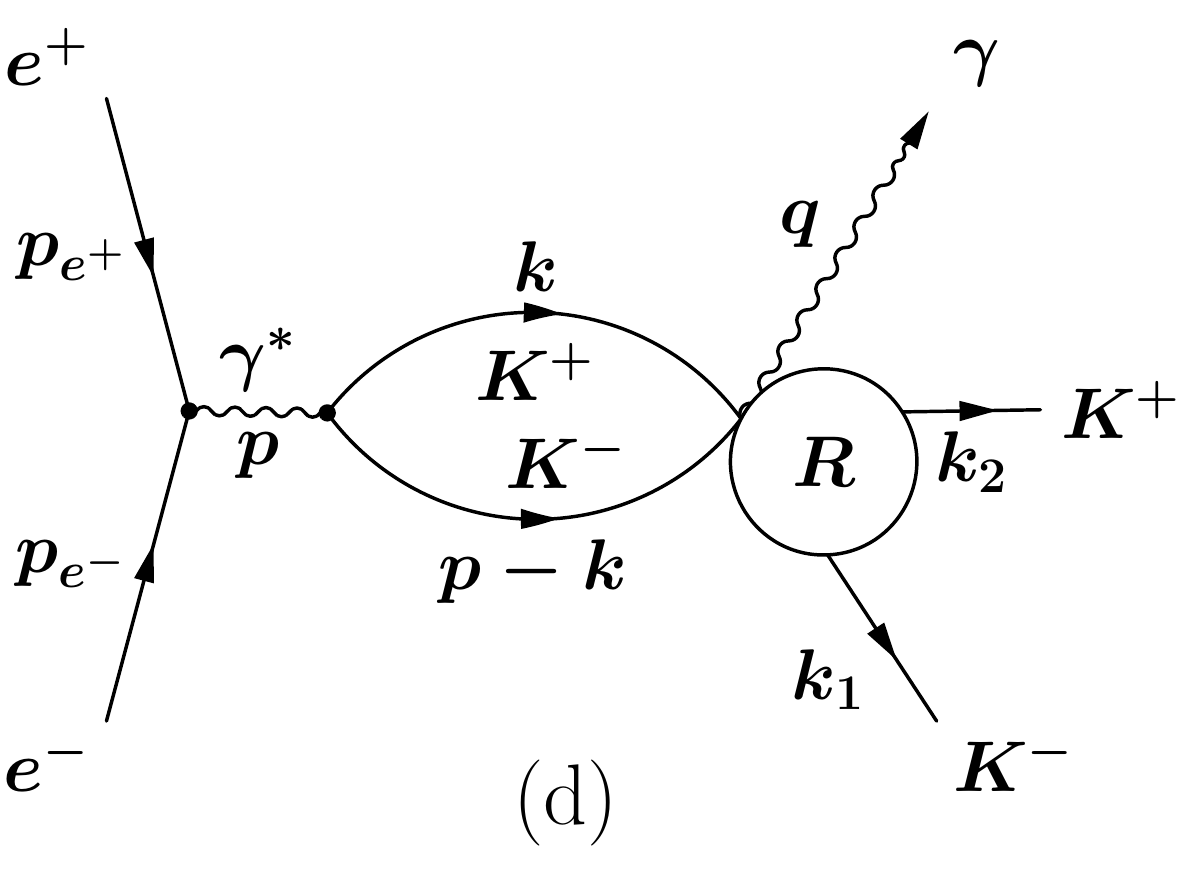}
\caption{Diagrams including the final-state $K^+ K^-$ interactions in the process $e^+ e^- \rightarrow K^+ K^- \gamma$.
$T$ denotes the $K^+K^-$ elastic scattering amplitude, $R$ stands for the difference of the $K^+K^-$ amplitudes present in Eq.~(\ref{A4}).
}
\label{fig-2}       
\end{figure}

In the FSR amplitude strong interactions between kaons are not yet included. However, it is possible to formulate a model of the final-state kaon interactions. It has been outlined in Ref.~\cite{LST}
and it is presented below in more detail. The corresponding diagrams are shown in Fig.~\ref{fig-2}.
The diagrams in Figs. 2(a,b,c) are directly connected to the diagrams present in Figs. 1(a,b,c), respectively. Here one includes a strong interaction between the $K^+K^-$ pairs in the final state.
The elastic $K^+ K^-$ scattering amplitude denoted by $T$ 
follows a kaon loop with a variable four-momentum $k$ over which one has to integrate.
The initial four-momenta of the positron $e^+$ and of the electron $e^-$ are denoted by $p_{e^+}$ and $p_{e^-}$, respectively.
The $K^-$, $K^+$ and photon four-momenta in the final state are labelled by $k_1$, $k_2$ and $q$, respectively.

The amplitudes for the $e^+ e^- \to K^+ K^- \gamma$ reaction,
corresponding to diagrams (a), (b), (c) in Fig.~\ref{fig-2} are given by:

\be
\label{A1}
A_1=2 i \int \frac{d^4 k}{(2 \pi)^4} \frac{J_{\nu} \epsilon^{\nu *} T(k)}{D(k) D(-k+p-q)},
\end{equation}

\be
\label{A2}
A_2=-4 i \int \frac{d^4 k}{(2 \pi)^4} \frac{J_{\mu} \epsilon^{\nu *} k_{\nu} (k_{\mu}+q_{\mu})T(k)}{D(k+q) D(k) D(-k+p-q)},
\end{equation}

\be
\label{A3}
A_3=-4 i \int \frac{d^4 k}{(2 \pi)^4} \frac{J_{\mu} \epsilon^{\nu *}~ (k_{\nu}-p_{\nu}) k_{\mu} T(k)}{D(p-k) D(k) D(-k+p-q)},
\end{equation}
where $D(k)=k^2-m_{\rm K}^2 + i \delta$, $\delta \rightarrow +0$, is the inverse of the kaon propagator, $m_{\rm K}$ is the charged kaon mass and $\epsilon^{\nu}$ is the photon polarization four-vector.
In the above expressions $p=p_{e^+}+p_{e^-}$, $J_{\mu}$ is defined as
\be
\label{J}
J_{\mu}=\frac{e^3}{s} F_K(s) \bar{v}(p_{e^+}) \gamma_{\mu} u(p_{e^-}),
\end{equation}
where $e$ is the electron charge, $s=(p_{e^+}+p_{e^-})^2$, $v$ and $u$ are the $e^+$ and $e^-$ bispinors,
respectively, $\gamma_{\mu}$ are the Dirac matrices and
$F_K(s)$ is the kaon electromagnetic form factor.
Appearance of the factor $e^3$ in Eq.~(4) is related to a presence of the three photon couplings which are most  easily seen in Figs.~2(b) and 2(c). The factor $1/s$ is the intermediate photon propagator. The virtual photon $\gamma^*$ couples to the $K^+K^-$ pair leading to a presence of the kaon form factor.  
The $K^+ K^-$ elastic scattering amplitude is given by
\be 
\label{Tk}
T(k)=\langle K^-(k_1) K^+(k_2)|\tilde{T}(m)|K^-(-k+p-q) K^+(k)\rangle,
\end{equation}
where $m^2=(k_1+k_2)^2$ is the square of the $K^+ K^-$ effective mass and
 $\tilde{T}(m)$ is the $K \bar K$ scattering operator.
The amplitude $T(k)$ depends not only on the $K^+$ four-momentum $k$ but also on the four-momenta $k_1$, $k_2$, $p$ and $q$ which satisfy the relation $k_1+k_2=p-q$. So $T(k)$ is a shorthand notation which underlines the dependence of the $K^+K^-$ off-shell amplitude $T$ on the kaon-loop momentum variable $k$ over which one has to integrate in Eqs.~(1-3). 
Let us denote by $T_S(k)$ the following sum of amplitudes:
\be 
\label{TS}
T_S(k)=T(k)+T(-k+p-q).
\end{equation}
Then one can rewrite Eqs.~(\ref{A1})-(\ref{A3}) as:
\be
\label{A1S} 
A_1=i \int \frac{d^4 k}{(2 \pi)^4} \frac{J_{\nu} \epsilon^{\nu *} T_S(k)}{D(k) D(-k+p-q)},
\end{equation}

\begin{equation}
\label{A23}
\begin{gathered}
A_2+A_3= -4 i J_{\mu} \epsilon^{\nu *}\\
 \times \int \frac{d^4 k}{(2 \pi)^4} \frac{k_{\nu} (k_{\mu}+q_{\mu})T_S(k)}{
D(k+q) D(k) D(-k+p-q)}.
\end{gathered}
\end{equation}

Oppositely to the sum of the FSR amplitudes shown in Fig.~1, the sum $S_3$ of the amplitudes  presented in Fig.~\ref{fig-2} (a), (b) and (c) is not gauge-invariant. 
This can be seen by the substitution
$\epsilon^{\nu} \to q^{\nu}$ into the sum $S_3\equiv A_1+A_2+A_3$ which after some algebra leads to the following result:
\be
\label{Agauge}
S_3(\epsilon^{\nu} \to q^{\nu})= 2 i \int \frac{d^4 k}{(2 \pi)^4} \frac{J \cdot k~[T(k-q)-T(k)]}{D(k) D(p-k)}.
\end{equation}
Thus, in order to satisfy the gauge invariance condition of the total amplitude: 
\be 
\label{Atot}
A\equiv A_1+A_2+A_3+A_4
\ee
we postulate the following form of the additional term $A_4$ which should be added to $S_3$ 
\begin{equation}
\label{A4}
\begin{gathered}
A_4=-2 i \int \frac{d^4 k}{(2 \pi)^4} \frac{J \cdot k~ \epsilon^* \cdot \tilde{k}}{D(k) D(p-k)}
\frac{[T(k-q)-T(k)]}{q\cdot \tilde{k}}.
\end{gathered}
\end{equation}
Here the four-vector $\tilde{k}=(0,\hat{\mathbf{k}})$ and the unit three-vector $\hat{\mathbf{k}}=\mathbf{k}/|\mathbf{k}|$. 

The integrals over the energy $k_0$ in Eqs.~(\ref{A1S}, \ref{A23}, \ref{A4}) can be done analytically in the $K^+ K^-$ center-of-mass frame and the results are:
\begin{align}
\label{A1k}
A_1=-\vec{J} \cdot \vec{\epsilon}~^*
\int \frac{d^3 k}{(2 \pi)^3}\frac{1}{2E_km} \left(\frac{T_1(\mathbf{k})}{m-2E_k}+\frac{T_2(\mathbf{k})}{m+2E_k}\right),\\
\label{A23k}
\nonumber
A_2+A_3=2 \vec{J}\cdot \vec{\epsilon}~^*
\int \frac{d^3 k}{(2 \pi)^3}
\left[ |\mathbf{k}|^2-(\mathbf{k}\cdot \mathbf{\hat{q}})^2 \right]\\ 
\times \left(\frac{T_1(\mathbf{k})}{M_1(\mathbf{k})}+\frac{T_2(\mathbf{k})}{M_2(\mathbf{k})}-\frac{T_3(\mathbf{k})}{M_3(\mathbf{k})}\right)
\end{align}
and
\be
\label{A4k}
\begin{gathered}
A_4=\vec{J}\cdot \vec{\epsilon}~^* ~ \int \frac{d^3 k}{(2 \pi)^3}
\frac{|\mathbf{k}|^2-(\mathbf{k}\cdot \mathbf{\hat{q}})^2}{\mathbf{k}\cdot \mathbf{q}}\\ 
\times \left(\frac{R_1(\mathbf{k})}{M_4(\mathbf{k})}+\frac{R_2(\mathbf{k})}{M_5(\mathbf{k})}\right).
\end{gathered}
\end{equation}

Here $\vec{\epsilon}$ and $\vec{J}$ are the three-component vectors 
of $\epsilon^{\nu}$ and $J_{\mu}$, respectively, and 
$\mathbf{\hat{q}}$ is the unit vector pointing in the direction of the emitted photon momentum.
In order to derive the formulae (\ref{A1k}) and (\ref{A23k}) one applies the relation:
\be
\label{Joteps}
J \cdot k~ \epsilon^*\cdot \tilde{k} = \frac{1}{2}~[~|\mathbf{k}|^2-(\mathbf{k}\cdot \mathbf{\hat{q}})^2]~\vec{J} \cdot \vec{\epsilon}~^*.
\end{equation}
The numerators in Eqs.~(\ref{A1k}), (\ref{A23k}) and (\ref{A4k}) are given by the following expressions:
\begin{widetext}
\begin{eqnarray} 
\label{T1}
T_1(\mathbf{k})=
\langle K^-(k_1) K^+(k_2) 
|\tilde{T}(m)|K^-(E_k,-\mathbf{k})K^+(m-E_k,\mathbf{k})
\rangle \\ \nonumber
+\langle K^-(k_1) K^+(k_2)|\tilde{T}(m)|K^-(m-E_k,-\mathbf{k})K^+(E_k,\mathbf{k})
\rangle, 
\end{eqnarray}
\begin{eqnarray}
\label{T2}
T_2(\mathbf{k})=\langle K^-(k_1) K^+(k_2)
|\tilde{T}(m)| K^-(m+E_k,-\mathbf{k})K^+(-E_k,\mathbf{k})\rangle \\ \nonumber
+ \langle K^-(k_1) K^+(k_2)
|\tilde{T}(m)|K^-(-E_k,-\mathbf{k})K^+(m+E_k,\mathbf{k})\rangle,
\end{eqnarray}
\begin{eqnarray}
\label{T3}
T_3(\mathbf{k})=\langle 
K^-(k_1) K^+(k_2) |\tilde{T}(m)|
K^-(m+E_{k+q}+\omega,-\mathbf{k})K^+(-E_{k+q}-\omega,\mathbf{k})
\rangle \\ \nonumber
+ \langle K^-(k_1) K^+(k_2)
|\tilde{T}(m)|K^-(-E_{k+q}-\omega,-\mathbf{k})K^+(m+E_{k+q}+\omega,\mathbf{k})\rangle ,
\end{eqnarray}
\begin{eqnarray}
\label{R1}
R_1(\mathbf{k})=\langle K^-(k_1) K^+(k_2) 
|\tilde{T}(m)|K^-(E_{k-q},-\mathbf{k}+\mathbf{q})K^+(m-E_{k-q},\mathbf{k}-\mathbf{q})\rangle \\ \nonumber
- \langle K^-(k_1) K^+(k_2)
|\tilde{T}(m)|K^-(E_{k-q}-\omega,-\mathbf{k})K^+(p_0-E_{k-q},\mathbf{k})\rangle ,
\end{eqnarray}
\begin{eqnarray}
\label{R2}
R_2(\mathbf{k})=\langle K^-(k_1) K^+(k_2)
|\tilde{T}(m)|K^-(E_k+m-\omega,-\mathbf{k}+\mathbf{q})K^+(-E_k-\omega,\mathbf{k}-\mathbf{q})\rangle \\ \nonumber
- \langle K^-(k_1) K^+(k_2) 
|\tilde{T}(m)|K^-(E_k+m,-\mathbf{k})K^+(-E_k,\mathbf{k})\rangle ,
\end{eqnarray}
\end{widetext}
where $E_{k\pm q} =\sqrt{\mathbf{(k\pm q)}^2+m_{\rm K}^2-i\delta}$, $\omega$ is the photon energy
and $E_k =\sqrt{\mathbf{k}^2+m_{\rm K}^2-i\delta}$.
Below we write expressions for the denominators in Eqs.~(\ref{A23k}) and ~(\ref{A4k}):
\begin{equation}
\label{M1}
M_1(\mathbf{k})=2 E_k m (m-2 E_k)(2 p_0 E_k- s +2\mathbf{k\cdot q)}, 
\end{equation}
\begin{equation} 
\label{M2}
M_2(\mathbf{k})=2 E_k m (m+2 E_k)(2 \omega E_k+2\mathbf{k\cdot q)}, 
\end{equation}
\begin{eqnarray}
\label{M3} 
\nonumber
M_3(\mathbf{k})=2 E_{k+q}(p_0^2+\omega ^2 + 2 p_0 E_{k+q} +
2 \mathbf{k \cdot q)} \\
\times (2 \omega ^2 + 2\omega E_{k+q}+2\mathbf{k\cdot q}),
\end{eqnarray}
\begin{equation}
\label{M4} 
M_4(\mathbf{k})=2 E_{k-q}(p_0^2+\omega ^2 - 2 p_0 E_{k-q} -
2 \mathbf{k \cdot q)},
\end{equation}
\begin{equation}
\label{M5}
M_5 (\mathbf{k})=2 E_k (2 p_0 E_k+ s+2\mathbf{k\cdot q)}.
\end{equation}
In the $K^+ K^-$ center-of-mass frame the energy $p_0$ equals to $m+\omega$, and the photon energy \mbox{$\omega=(s-m^2)/(2 m)$}. 
The formulae ~(\ref{A1k}-\ref{A4k}) and ~(\ref{T1}-\ref{M5}) constitute a full set of expressions for the general form of the reaction amplitude $A$ for the process $e^+ e^- \to K^+ K^- \gamma$. In Sec.~\ref{other} one can find an extension of this formalism to processes with other pseudoscalar meson pairs in the final state.

Now we can examine some approximations to the reaction amplitude $A$.
Let us denote by $T_{K^+ K^-}(m)=\langle K^-(k_1) K^+(k_2)|\tilde{T}(m)|K^-(k_1) K^+(k_2)\rangle$ the on-shell $K^+ K^-$ amplitude.
In the $K^+ K^-$ center-of-mass frame $k_1=(m/2,-\mathbf{k_f})$,
$k_2=(m/2,\mathbf{k_f})$  and $k_f=\sqrt{m^2/4-m_{\rm K}^2}$ is the kaon momentum in the final-state.
As seen from Eq.~(\ref{T1}) $\mathbf{k}$ is equal to the half of the difference between the $K^+$ and $K^-$ momenta, so it is the relative momentum of the kaon pair in their center-of-mass frame.
For the value $E_k=m/2$ the amplitude $T_1(\mathbf{k})$ is equal to the doubled on-shell $K^+ K^-$ amplitude.
Therefore we can assume that $T_1(k)$ is related to $T_{K^+ K^-}(m)$ as follows:
\be
\label{Ton}
T_1(\mathbf{k})\approx 2 g(k) T_{K^+ K^-}(m),
\end{equation}
where $g(k)$ as a real function of $k\equiv|\mathbf{k}|$ takes into account
the off-shell character of $T_1(\mathbf{k})$\footnote{For some specific $K^+K^-$ amplitudes Eq.~(\ref{Ton}) is exact. This is a case of separable interactions discussed in Sec.~\ref{KKamplitudes}.}.  
We note here that the function $g(k)$ satisfies the condition $g(k_f)=1$.

If $E_k=m/2$ then the denominator $M_1(\mathbf{k})=0$, so the first terms in parentheses in Eqs.~(\ref{A1k}) and (\ref{A23k}) have a pole. 
Thus, at not too large values of $|\mathbf{k}|$ one expects a dominance of these terms over the other ones which depend on $T_2(\mathbf{k})$ or $T_3(\mathbf{k})$
in Eqs.~(\ref{A1k}) and (\ref{A23k}). So, omitting temporarily the terms with $T_2(\mathbf{k})$ and $T_3(\mathbf{k})$,  one derives the approximate sum of the first three amplitudes of our reaction in the following form:

\be
\label{Tap} 
A_1+A_2+A_3 \approx ~\vec{J} \cdot \vec{\epsilon}~^*~T_{K^+ K^-}(m)~I(m),
\end{equation}
where the integral $I(m)$ reads
\begin{eqnarray}
\label{I1}
\nonumber
I(m)= -2
~\int \frac{d^3 k}{(2 \pi)^3}
\frac{g(k)}{2E_km(m-2E_k)}\\
\times \left[1-2~\frac{|\mathbf{k}|^2-(\mathbf{k\cdot \hat{q})^2}}{2 p_0 E_k- s +2\mathbf{k\cdot q)}}\right].
\end{eqnarray}
One can expect that the function $g(k)$ decreases for the momenta $k$ going to infinity. In order to make the integral in Eq.~(\ref{I1}) finite, the function $g(k)$ should decrease at large $k$ steeper than $1/k^2$. If this is not a case for a particular model of the $K^+ K^-$ amplitude, then one has to replace $g(k)$ by
another function $\tilde{g}(k)$ to warrant an integral convergence. One can also choose an upper limit cut-off $k_{\rm{cut}}$ parameter for the integral over $k$.

There is an alternative form of the approximate sum of the amplitudes $A_1$, $A_2$ and $A_3$. 
One can notice that the relative kaon momenta in the expressions~(\ref{T2}) and ~(\ref{T3}) for the amplitudes $T_2(\mathbf{k})$ and $T_3(\mathbf{k})$ are the same as the momentum $\mathbf{k}$ in $T_1(\mathbf{k})$. So, similarly to Eq.~(\ref{Ton}) the following approximation can be chosen:
\be
\label{T2T3}
T_2(\mathbf{k}) \approx T_3(\mathbf{k}) \approx 2 g(k) T_{K^+ K^-}(m).
\ee
Thus one can write an alternative form of the amplitude sum:
\be
\label{Tap2} 
A_1+A_2+A_3 \approx  ~\vec{J} \cdot \vec{\epsilon}~^*~T_{K^+ K^-}(m)~I_{\rm r}(m),
\end{equation}
where
\begin{eqnarray}
\label{Ir}
\nonumber
I_{\rm r}(m)=
\;\; -2 \int \frac{d^3 k}{(2 \pi)^3}g(k)
 \{
\frac{1}{ E_k (m^2-4 E_k^2)} \\
\!\!\!\!\!
 - 2\left[|\mathbf{k}|^2-(\mathbf{k\cdot \hat{q})^2}\right]
\left[\frac{1}{M_1(k)}+\frac{1}{M_2(k)}-\frac{1}{M_3(k)}\right]  \}.
\end{eqnarray}
Due to a presence of the pole at $m^2=4E_k^2$ in the integrand of 
$I_{\rm r}(m)$, one can call it the "relativistic" version of $I(m)$.
Both the integrals $I$ and $I_{\rm r}$ will be called the kaon-loop functions.
It should be mentioned here that the imaginary parts of these functions are identical and only the real parts differ.
The equality of $\mathfrak{Im} I(m)$ and $\mathfrak{Im} I_{\rm r}(m)$ follows from the structure of the denominators in Eqs.~(\ref{A1k}) and~(\ref{A23k}). Only two poles
of $M_1(\mathbf{k})$ give contributions to the imaginary parts, namely the first one at $m=2 E_k$ and the second at $2 p_0 E_k- s +2
 \mathbf{k \cdot q}=0$.
The pole of the amplitude $A_1$ coincides with the first one.

For the real part of the second kaon-loop function $I_{\rm r}(m)$ one gets much better convergence of the integrand than for $I(m)$. If one makes an expansion of the corresponding integrand in series of the photon energy $\omega$ then  due to specific cancellations between the three terms in Eq.~(\ref{Ir}) at high $k$ momenta one gets a proportionality to $1/k^3$ while the integrand of $I(m)$ without $g(k)$ is directly proportional to $k$.
This happens when we take into account the part of the integrand linearly proportional to $\omega$. The term proportional to $\omega^2$ is proportional to $1/k^5$, so the convergence of $\mathfrak{Re}~ I_{\rm r}(m)$ at the infinite $k$ is even better. A limit $\omega$ going to zero will be discussed in Subsection ~\ref{limitqzero}.
 
After this general discussion we can pass to examination of the effective mass dependence of the kaon-loop functions.
It is possible to perform analytically two integrations over the angles of the vector $\mathbf{k}$ in the expression for the function $I(m)$ in Eq.~(\ref{I1}). The results for the real and imaginary parts are:
\begin{eqnarray}
\label{ReI}
\nonumber
\mathfrak{Re}~ I(m)= -\frac{1}{2\pi^2} ~{\rm P} \int_0^{\infty} dk  
~\frac{k^2~g(k)}{2E_km(m-2E_k)}\\
 \times \left \{2-\frac{1}{\omega k}\left [2 y k + (k^2 - y^2) \ln \left |\frac{y+k}{y-k}\right|\right] \right\},
\end{eqnarray}
where $\rm P$ standing before the integral symbol denotes the principal value part, 
\be
\label{y}
y=\frac{1}{\omega}\left(p_0~E_k-\frac{s}{2}\right),
\end{equation}
\be 
\label{ImI}
\mathfrak{Im}~ I(m)=\mathfrak{Im}~ I_a(m)+ \mathfrak{Im}~I_b(m),
\end{equation}
\begin{eqnarray} 
\label{ImIa}
\begin{gathered}
\mathfrak{Im}~ I_a(m)=
\frac{k_f}{8 \pi m}\\ 
\times \left[2+\frac{1}{\omega k_f}\left(m~ k_f + m_{\rm K}^2 \ln\frac{m-2k_f}{m+2k_f}\right)\right]
g(k_f)
\end{gathered}
\end{eqnarray}
and
\be 
\label{ImIb}
\mathfrak{Im}~ I_b(m)=
\frac{1}{2\pi\omega}\int_{k'}^{k''} dk 
~\frac{k^3~g(k)(1-y^2/k^2)}{2E_km(m-2E_k)}.
\end{equation}
In Eq.~(\ref{ImIb}) $k^{'}=(p_0 v_L -\omega)/2$, $k^{''}=(p_0 v_L +\omega)/2$ and
$v_L=\sqrt{1-4 m_{\rm K}^2/s}$ is the kaon velocity in the $e^+e^-$ center-of-mass frame at $m^2=s$. The first term in the numerator of this equation corresponds to a pole contribution at $E_k=m/2$ in the first denominator of the integrand in Eq.~(\ref{I1}) and the second term, proportional to $y^2/k^2$, is related to a pole of the second denominator. In the latter case the position of the pole depends on the angle between the vectors $\bf{k}$
and $\bf{q}$ which leads to an integration over a range of $k$ between the limiting values $k^{'}$ and $k^{''}$. It can be checked that both $k^{'}$ and $k^{''}$ values are slightly larger than $k_f$.
 If the function $g(k)$ is equal to 1 in the $k$ range below $k^{''}$, then one can perform integral in Eq.~(\ref{ImIb}), so the second term of $\mathfrak{Im}~ I(m)$ reads:
\be
\label{ImIbis}
\mathfrak{Im}~ I_b(m)= -\frac{1}{8\pi \omega m}\left(\frac{1}{2}v_L s + m_{\rm K}^2 
\ln\frac{1-v_L}{1+v_L}\right).
\end{equation}
In addition, if $m$ is approaching its maximum value of $\sqrt{s}$, or equivalently in the limit of vanishing photon energy $\omega$, the function $ \mathfrak{Im}~I(m)$ goes to zero since $\mathfrak{Im}~I_a(m)=-\mathfrak{Im}~I_b(m)$. 

Let us now pass to a discussion of the properties of the amplitude $A_4$ given by Eq.~(\ref{A4k}).
Using the similar assumption as that leading to Eq.~(\ref{Ton}) one can approximate $R_1$ and $R_2$ as
\begin{equation}
\label{R1ap} 
R_1 \approx R_2 \approx [g(|\mathbf{k - q}|) - g(|\mathbf{k}|)]~ T_{K^+ K^-}(m).
\end{equation}
At small values of $|\mathbf{q}|$ with respect to $|\mathbf{k}|$, $R_1$ and $R_2$ can be further approximated as
\begin{equation}
\label{R1der}
R_1 \approx R_2 \approx - \mathbf{q} \cdot \hat{\mathbf{k}}~ g'(|\mathbf{k}|)~T_{K^+ K^-}(m),
\end{equation}
where $g'(|\mathbf{k}|)$ is the derivative of the function $g(|\mathbf{k}|)$ responsible for the off-shell character of the $K\bar{K}$ scattering.
In this way the amplitude $A_4(m)$ from Eq.~(\ref{A4k}) reads
\be
\label{A4k3}
A_4(m)\approx \vec{J}\cdot \vec{\epsilon}~^* ~T_{K^+ K^-}(m) I_4(m),
\ee
where
\be 
\label{I4}
I_4(m)=- \int \frac{d^3 k}{(2 \pi)^3}g'(|\mathbf{k}|)|\mathbf{k}|[1-(\mathbf{\hat{k}} \cdot \mathbf{\hat{q}})^2]
[\frac{1}{M_4(\mathbf{k})}+\frac{1}{M_5(\mathbf{k})}].
\ee
For the imaginary part of $I_4(m)$ the integration over the two angles of the vector $\mathbf{k}$ can be performed and the result is
\be
\label{ImI4k}
\mathfrak{Im}~ I_4(m)= 
\frac{1}{16 \pi \omega} \int_{k'}^{k''} dk ~\frac{g'(k)(k^2-y^2)}{E_k}. 
\ee
Here the variable $y$ is defined by Eq.~(\ref{y}) and the integral limits $k^{'}$ and $k^{''}$ are defined just below Eq.~(\ref{ImIb}).

The amplitude $A_4(m)$ from Eq.~(\ref{A4k3}) has to be added to the sum of the amplitudes $A_1$, $A_2$ and $A_3$ in 
Eqs.~(\ref{Tap}) or (\ref{Tap2}). Then the total reaction amplitude $A$ is formed (Eq.~\ref{Atot}).

\subsection{Limit $\omega \to 0$}
\label{limitqzero}

In the limit of vanishing photon energy $\omega \equiv |\mathbf{q}|\to 0$ and at $s$ equal to the square of the $\phi(1020)$ meson mass
$m_{\phi}$ one gets the following relation for the imaginary part of $I_4(m)$
given by Eq.~(\ref{ImI4k}):
\begin{equation}
\label{ImI4zero}
\mathfrak{Im}~ I_4(m_{\phi})= 
\frac{g'(|\mathbf{k_{\phi}|)k_{\phi}^2}}{12 \pi m_{\phi}},
\end{equation}
where the momentum $k_{\phi}=\sqrt{m_{\phi}^2/4-m_{\rm K}^2}$. 
It is interesting to find a close relation of this formula to the imaginary part of the loop function $I(m)$ (Eq.~(\ref{I1})), calculated in the limit $m \to m_{\phi}$, which is equivalent to the limit $\omega \to 0 $:
\begin{equation}
\label{IrA}
 \mathfrak{Im}~ I_4(m_{\phi}) = - \mathfrak{Im}~ I(m_{\phi}).
\end{equation}

Below we show that the above relation is valid also for the real parts of the above functions which leads to a relation between the four amplitudes at $\omega=0$: 
\begin{equation}
\label{rownIm}
A_4(m_{\phi})= -[A_1(m_{\phi})+A_2(m_{\phi})+A_3(m_{\phi})].
\end{equation} 
Let us sketch a derivation of this formula.
Going back to Eq.~(\ref{A4}) we use Eqs.~(\ref{R1},\ref{R2},\ref{R1ap},\ref{R1der}) to get the approximate expression for the amplitude $A_4$ valid for small $\omega$:
\begin{eqnarray}
\label{A4apr}
\nonumber
A_4(m) \approx -2 i ~T_{K^+ K^-}(m)\\
\times \int \frac{d^4 k}{(2 \pi)^4} \frac{J \cdot k~ \epsilon \cdot \tilde{k}}{D(k) D(p-k)}
g'(|\mathbf{k}|).
\end{eqnarray}
Knowing that in the $K^+ K^-$ center-of-mass-frame the momenta $\mathbf{p}$ and $\mathbf{q}$ are equal we can assume as in Eq.~(\ref{Ton}) that the sum of the amplitudes $T_S(k)$ in Eq.~(\ref{TS})
can be expressed in terms of the function $g(|\mathbf{k}|)$ of the relative kaon momentum as 
$T_S(k)\approx 2 g(|\mathbf{k}|) ~T_{K^+ K^-}(m)$. Then, from Eqs.~(\ref{A1S}) and ~(\ref{A23}),
one gets the following amplitudes in the limit $\omega \to 0$:
\begin{eqnarray}
\label{A1z}
\nonumber
A_1(m_{\phi})\approx J_{\nu} \epsilon^{\nu *}~2 i ~T_{K^+ K^-}(m_{\phi})\\
\times \int \frac{d^4 k}{(2 \pi)^4}\frac{g(|\mathbf{k}|)}{D(k) D(p-k)},\\
\label{A2+3z}
\nonumber
A_2(m_{\phi})+A_3(m_{\phi}) \approx -J_{\mu} \epsilon^{\nu *}~ 8 i ~T_{K^+ K^-}(m_{\phi})\\
\times \int \frac{d^4 k}{(2 \pi)^4} \frac{k_{\nu} k_{\mu}g(|\mathbf{k}|)}{[D(k)]^2 D(p-k)}.
\end{eqnarray}
Next, after an integration over energy and in the next step by integrating the amplitude $A_4(m)$ in Eq.~(\ref{A4apr}) by parts over $|\mathbf{k}|$, one finds that 
Eq.~(\ref{rownIm}) is satisfied.
To get this result one has to assume that the function $g(|\mathbf{k}|)$ tends to 0 when $|\mathbf{k}|$ goes to infinity. 
In consequence, the full reaction amplitude $A\equiv A_1+A_2+A_3+A_4$ vanishes in the limit $\omega \to 0$. 
This is a consequence of the gauge invariance of the total reaction amplitude $A$.

In Appendix A we show that the amplitude $A_4(m)$ depends weakly on the variable $m$ or $\omega$, so to a very good approximation
$A_4(m)\approx A_4(m_{\phi})$.

Recalling the relations given in Eqs.~(\ref{Atot}), (\ref{Tap}) and ~(\ref{rownIm}) for $\sqrt{s}=m_{\phi}$ we can write the following expression for the reaction amplitude $A(m)$:
\begin{equation}
\label{Atot2}
A(m)= ~\vec{J} \cdot \vec{\epsilon}~^*~T_{K^+ K^-}(m)~[I(m)-I(m_{\phi})],
\end{equation}
where the loop function $I(m)$ is given by Eq.~(\ref{I1}) or by  Eq.~(\ref{Ir}).
We stress here that the imaginary parts of $I(m)$ and $I_{\rm r}(m)$ are equal and the corresponding formulae are given by Eqs.~(\ref{ImI}-\ref{ImIb}).
The real part of the function $I(m)$ is seen in Eq.~(\ref{ReI})
and the formulae for the real part of the kaon-loop function
$I_{\rm r}$ are written in Appendix B.
The integrand of the real part of the function $I(m)$ is simpler than the corresponding integrand of $\mathfrak{Re}~ I_{\rm r}(m)$ but on the other hand the convergence at high $k$ is much better for the latter function.  

\subsection{Comparison of the present model with other approaches}
\label{otherappr}

We may see some similarity of the formulae presented above with the expressions for the amplitudes of the radiative $\phi$ meson decays derived in Ref.~\cite{Close}. 
In particular, Eq.~(\ref{A4apr}) can be compared with Eq.~(4.24) of Ref.~\cite{Close} if we replace the current $J$ by the $\phi$ meson polarization vector $\epsilon_{\phi}$ and the function $g(k)$ by the function $\phi(k)$. 
One has also to multiply the amplitude $A_4$ by $i$. The same multiplication factor $i$ should be  
applied to the amplitudes $A_1$, $A_2$ and $A_3$ in order to make a comparison with Eqs.
(4.21), (4.22) and (4.23) of Ref.~\cite{Close}.
Thus the model of Close, Isgur and Kumano for the $\phi$ radiative decay amplitudes 
is a special case of the present model in which the reaction amplitudes are given by Eqs.~(\ref{A1})-(\ref{A3}) and~(\ref{A4}), and the photon momentum $q$ is small (soft photon limit). 

One can notice a difference in the normalization of the functions $g(k)$
and $\phi(k)$. 
The latter function is defined by Eq. (4.14) of Ref.~\cite{Close} as follows:
\begin{equation}
\label{fik}
\phi(k)=\frac{\mu^4}{(k^2+\mu^2)^2},
\end{equation}
where the parameter $\mu=141$ MeV.
The function $g(k)$ has to be normalized to 1 at the final $K^+K^-$
relative momentum $k=k_f$ while $\phi(k)=1$ at $k=0$, so the function $g(k)$ related to $\phi(k)$ should be defined as
\begin{equation}
\label{giek}
g(k)=\frac{(k_f^2+\mu^2)^2}{(k^2+\mu^2)^2}.
\end{equation}
As we shall see later, the normalization condition $g(k_f)=1$, instead of $g(0)=1$, has an important influence on the values of the kaon-loop function when the range parameter $\mu$ is relatively small.

In the model of Achasov, Gubin and Shevchenko~\cite{KL} the $\phi$ decay amplitude
is regularized by making a subtraction at the photon energy $\omega=0$. So, effectively one can state that in their model the amplitude $A_4(m)=A_4(m_{\phi})=-[A_1(m_{\phi})+A_2(m_{\phi})+A_3(m_{\phi})]$. Thus the above approach could also be treated as a particular version of the model introduced in Subsection \ref{model derivation}.

At $\sqrt s$ close to 1 GeV  the $K^+K^-$ scattering amplitude is usually taken in the following resonant form
\begin{equation}
\label{res}
T_{\rm {res}}(m)=\frac{(g_{R K^+K^-})^2}{D_R(m)},
\end{equation}
where $g_{R K^+K^-}$ is the scalar resonance coupling constant to the $K^+K^-$ pair
and $D_R(m)$ is the inverse of the scalar meson propagator. 
Here $R$ denotes the scalar mesons $a_0(980)$ or $f_0(980)$. 
Formally, this case is  a
point-like version of the $K^+K^-$ scattering amplitude $T_{K^+ K^-}(m)$, since here the function $g(k)\equiv 1$,
$g'(k)\equiv 0$ and the amplitude $A_4(m)\equiv 0$.
The resonant $K^+K^-$ scattering amplitude has been used in Ref.~\cite{Aczasow2001}.
It has been multiplied by the kaon-loop function taken from Ref.~\cite{Ivan}. The real part form of the latter function has been obtained applying twicely subtracted dispersion relations constrained by gauge invariance. The kaon-loop function, constructed in that way, also vanishes at $\omega \to 0$.


\section{$K^+ K^-$ scattering amplitude}
\label{KKamplitudes}

The elastic on-shell scattering amplitude $T_{K^+ K^-}(m)$ is normalized using the following relation to the elastic $S$-matrix element $S_{K^+ K^-}$: 
\begin{equation} 
\label{norm}
T_{K^+ K^-}(m)=\frac{4\pi m}{i k_f}(S_{K^+ K^-}-1).
\end{equation}
Like the function $I(m)$ the above amplitude is dimensionless. 

The $K^+ K^-$ $S$-wave state can be decomposed into two isospin states corresponding to isospin $I=0$
or isospin $I=1$:
\begin{equation}
\label{isospin}
|K^+K^-\rangle =\frac{1}{\sqrt{2}}(|I=0\rangle +~|I=1\rangle).
\end{equation}
If one assumes isospin symmetry conservation in the $K^+K^-$ interaction, then the strong
elastic scattering amplitude $T_{K^+ K^-}(m)$ can be written as a linear combination of two isospin amplitudes $t_0(m)$ and $t_1(m)$:
\begin{equation}
\label{TI}
T_{K^+ K^-}(m)=\frac{1}{2}[t_0(m)+t_1(m)].
\end{equation}
The amplitudes $t_0$ and $t_1$ are the elastic transition amplitudes between the isospin 0 and 1 states, respectively.
Similarly, the $S_{K^+ K^-}$ matrix element is related to two $K \bar K$ elastic $S$-matrix  elements labelled by isospin 0 or 1:
\begin{equation}
\label{SI}
S_{K^+ K^-}=\frac{1}{2}(S_0+S_1).
\end{equation}
If the isospin symmetry is not conserved, then one can consider additional contributions to $T_{K^+ K^-}(m)$ or to $S_{K^+ K^-}$.

It is convenient to express the complex functions $S_I$, $I=0,1$, in terms of the real phase shifts $\delta_I$ and inelasticities $\eta_I$:
\begin{equation}
\label{deli} 
S_I=\eta_I e^{2i\delta_I}.
\end{equation}
The functions $\delta_I$ and $\eta_I$ depend on the effective mass $m$ of the $K \bar K$ system and near the $K \bar K$ threshold they can be   developed into series depending on the kaon momentum evaluated in the $K^+ K^-$ center-of-mass frame. Alternatively, one can make an effective range expansion of the scattering amplitude $T_{K^+ K^-}(m)$. As shown
in Ref.~\cite{LL}, this can be done also in presence of the poles corresponding to the scalar resonances $f_0(980)$ or $a_0(980)$ located near the kaon-kaon threshold.  

    There exist many models of kaon-kaon interactions.
We do not intend to review all of them, however we shall mention here a model of separable potentials which has been successfully used in scalar meson spectroscopy (see, for example, Refs.~\cite{KLL,Furman,LL96}).
Separable pion-pion potentials have been used in Ref.~\cite{Markushin}.
 Some kaon-kaon amplitudes with parameters fitted to experimental data will be used in next sections in numerical calculations of the cross-sections for the $e^+e^- \to K \bar{K} \gamma$ reactions. Below we give a few equations
specific for the separable interactions.

The simplest rank-one-potential of the $K \bar{K}$ interaction $V$ written in the momentum space has the form:
\begin{equation}
\label{Vsep}
\langle  \mathbf{k_f}|V| \mathbf{k_i} \rangle=\lambda~G(k_f)~ G(k_i),   
\end{equation}
where $\lambda$ is the potential strength constant, $\mathbf{k_i}$, $\mathbf{k_f}$ are
the initial and final state relative kaon momenta in the kaon-kaon center-of-mass frame,
$k_i$ and $k_f$ are their moduli
and $G(k)$ is the vertex form factor. 
The Yamaguchi form factor~\cite{Yamaguchi} reads:
\begin{equation}
\label{Yama}
G(k)=\sqrt{\frac{4 \pi}{m_{\rm K}}}\frac{1}{k^2+\beta^2},
\end{equation}
where $\beta$ is the form-factor range parameter.
For the separable potential the scattering amplitude $T_{\rm {sep}}$ can be obtained from the Lippmann-Schwinger equation in the factorizable form:
\begin{equation}
\label{Tsep}
\langle  \mathbf{k_f}|T_{\rm {sep}}| \mathbf{k_i} \rangle= G(k_f) ~\tau(m)~G(k_i),
\end{equation}
where $\tau(m)$ is the $K \bar{K}$ effective mass dependent function. In the lowest order of the $K \bar{K}$ interaction $\tau(m)$ equals to the coupling constant $\lambda$ but if the
$K \bar{K}$ interaction is strong enough the function $\tau(m)$ may acquire a resonant character. 
This function has a pole in the complex effective mass $m$-plane  which can be attributed to the scalar $S$-wave resonance.
However, it can have also a non-resonant part, so in general it cannot be reduced just to a simple Breit-Wigner representation of the scalar resonance. 

One can notice that the resonant form of the amplitude given by Eq.~(\ref{res}) can be interpreted as a special case of the amplitude derived for the separable potential meson-meson interactions (Eq.~\ref{Tsep}).
In this case the form factors $G(k)$ are functions which take into account the interactions of kaons treated as extended objects. In 
Eq.~(\ref{res}) the coupling constant $g_{R K^+K^-}$ is independent of the kaon momentum, so in this case the kaons are treated as point-like objects.
 
For the separable potentials it is easy to get the function $g(k)$ introduced in Eq.~(\ref{Ton}) in order to describe the off-shell dependence of the $K \bar{K}$ amplitudes. For the on-shell scattering the initial and final kaon momenta are equal, $|\mathbf{k_i}|= \mathbf{k_f}|$, so the corresponding on-shell amplitude is proportional to the square of $G(k_f)$ 
and the function $g(k_i)$ is a simple ratio of the form factors:
\begin{equation}
\label{gsep}
g(k_i)=\frac{G(k_i)}{G(k_f)}.
\end{equation}
For the Yamaguchi form factor from Eq.~(\ref{Yama}) with $k=k_i$ this function equals to
\begin{equation}
\label{gYama}
g(k)=\frac{k_f^2+\beta^2}{k^2+\beta^2}.
\end{equation}
As discussed earlier in the text below Eq.~(\ref{Ton}), for this particular
form factor the amplitude integral $I(m)$ is divergent, so in numerical calculations we use a cut-off limit $k_{\rm{cut}}$. 
As in Ref.~\cite{Oller} we take $k_{\rm{cut}}=1$ GeV. 

Before coming to numerical calculations of the reaction cross sections one needs to determine the $K^+K^-$ amplitudes in two isospin states. 
In the present application the separable meson-meson potentials are used for both isospin channels. The parameters of the separable potentials have been obtained from fits to the available data in meson channels coupled to the $K^+K^-$ states.

For isospin zero we shall use the results obtained from the three-channel model of Ref.~\cite{KLL} (fit A). 
This model has been constructed with an experimental input on the two-pion, two-kaon and four-pion production on hydrogen targets (see, for example, Ref.~\cite{kam2}).
In the fit to data the pole corresponding to the $f_0(980)$ resonance has been found with the mass equal to 989 MeV and the width of 62 MeV. 
The range parameter of the isospin zero $K \bar{K}$ potential has been obtained as $\beta\approx1.5$ GeV.
In fits leading to a set of separable potential parameters obtained in Ref.~\cite{KLL}, the kaon mass has been taken as an average of the charged and neutral kaon masses $m_{\rm {av}}$. 
Since presently we need to distinguish the $K^+ K^-$ and $K^0 \bar{K^0}$ thresholds,
in numerical calculations of the amplitude $t_0(m)$ in Eq.~(\ref{TI})
we have made a shift of the mass $m$ by about 2 MeV by changing the argument $m$ into $m~ (m_{\rm {av}}/m_{\rm K})$. 
The separable potential parameters of the amplitude $t_1(m)$ have been directly calculated for the charged kaon mass.

For isospin one we take amplitudes obtained in Ref.~\cite{Furman}.
Here the pion-eta and kaon-antikaon coupled-channel amplitudes have been calculated using the relevant data on the meson production including the Crystal Barrel Collaboration results from the proton-antiproton annihilation.
The position of the $a_0(980)$ resonance on the $(-~+)$ sheet has been fitted in~\cite{Furman} at the mass of
1005 MeV and the width of 49 MeV.
In this case the value $\beta \approx 21.8$ GeV for the isospin one range parameter has been obtained.

At the end of this chapter we give a few remarks about some specific features of the $K\bar{K}$ amplitudes in relation to the phase shifts and inelasticity parameters in two isospin channels.
For isospin zero and near the $K^+K^-$ threshold a rapid decrease of the amplitude modulus exists. It is related to a presence of the scalar-isoscalar resonance $f_0(980)$. 
Its influence leads to a strong decrease of the $K \bar{K}$ phase shifts $\delta_0$ as well as to the steep behaviour of the inelasticity $\eta_0$ as a function of $m$ near the $K \bar{K}$ threshold (see 
Figs.~2 and 3 in Ref.~\cite{KLL}).
For isospin one we do not observe such a
strong decrease of phase shifts, although the scalar-isovector resonance $a_0(980)$ 
is present as a pole in the $K^+K^-$ scattering amplitude. As shown in Fig. 3 of Ref.~\cite{FL}, a more smooth behaviour of $|T_{K \bar{K}}(m)|$ for isospin one in comparison with 
the isospin zero case is related to small values of the corresponding 
$K \bar{K}$ phase shifts. 

\section{Differential cross-sections, angular distributions and the branching fraction for the  decay $\phi(1020) \to K^+ K^- \gamma$}
\label{Diff}
The reaction amplitude $A$ given by Eq.~(\ref{Atot})
depends on the spin projections (helicities) of the initial electrons, positrons and the final photons. These helicities are labelled by $\lambda_{e^-}$, $\lambda_{e^+}$ and
by $\lambda_{\gamma}$, respectively. In general, there are eight helicity dependent amplitudes $A_{\lambda_{e^-}, ~\lambda_{e^+},~\lambda_{\gamma}}$
since the electron or positron helicities can be equal to $+1/2$ or $-1/2$ and the photon helicities can take values $+1$ or $-1$. If the initial beams are unpolarized and the photon polarisation is not measured, then one has to average the modulus of the amplitude squared over the initial $e^+$ and $e^-$ helicities and sum over the photon helicities:
\begin{equation} 
\label{av}
|\mathcal{M}|^2=\frac{1}{4} \sum_{\lambda_{e^-}, \lambda_{e^+},\lambda_{\gamma}}
|A_{\lambda_{e^-},~\lambda_{e^+},~\lambda_{\gamma}}|^2.
\end{equation}
The differential cross-section for the reaction $e^+ e^- \to K^+ K^- \gamma$
is proportional to the above sum over the particle helicities:
\begin{equation}
\label{dsig}
d\sigma=\frac{(2\pi)^4}{2\sqrt{s(s-4m_e^2)}}~|\mathcal{M}|^2~ d\Phi_3,
\end{equation}
where $m_e$ is the electron mass and $\Phi_3$ is the phase space of the 
three-body final state consisting of $K^+$, $K^-$ and $\gamma$.

The final-state phase space $d\Phi_3$ can be written as
\begin{equation}
\label{phi3}
d\Phi_3=\frac{1}{(2\pi)^9}\frac{k_f ~\omega_l}{8\sqrt{s}}~d\Omega_1 ~d\Omega_{\gamma}~ dm,
\end{equation}
where $\omega_l=(s-m^2)/(2\sqrt{s})$ is the final photon energy in the 
$e^+e^-$ center-of-mass frame, $\Omega_1$ is the $K^-$ solid angle in the 
$K^+K^-$ center-of-mass frame and $\Omega_{\gamma}$ is the photon solid angle in the $e^+e^-$ center-of-mass frame.

Taking into account properties of the electron and positron bispinors ($u$ and $v$ in Eq.~(\ref{J})) which satisfy Dirac's equations, one can derive the following result:
\begin{eqnarray}
\label{suma} 
\begin{gathered}
|\mathcal{M}|^2=\left(\frac{e^3}{s}\right)^2 |F_K(s)|^2~|T_{K^+ K^-}(m)|^2~
|I(m)-I(m_{\phi})|^2\\
\times \left [s\frac{(p_{e^+} \cdot q)^2+(p_{e^-} \cdot q)^2}{(q \cdot p)^2}+2 m_e^2 \right ].
\end{gathered}
\end{eqnarray} 
If we denote by $\theta_{\gamma}$ the angle between the photon and electron momenta in the $e^+ e^-$ center-of-mass frame, then 
$|\mathcal{M}|^2$ can be written as:
\begin{eqnarray}
\label{sumee}
\begin{gathered}
|\mathcal{M}|^2=\left(\frac{e^3}{s}\right)^2 |F_K(s)|^2~|T_{K^+ K^-}(m)|^2~
|I(m)-I(m_{\phi})|^2\\
\times \left [\frac{1}{2}s(1+\cos^2\theta_{\gamma})+2 m_e^2 (1-\cos^2\theta_{\gamma})\right].
\end{gathered}
\end{eqnarray}
If the $e^+ e^-$ energy in the center-of-mass frame is close to the $\phi(1020)$ meson mass $m_{\phi}$ the second term in parentheses of Eq.~(\ref{sumee}) can be neglected so the photon angular distribution in the $e^+e^-$ center-of-mass frame is proportional to $(1+\cos^2\theta_{\gamma})$. In the same frame the angular kaon 
distributions are constant as the kaons are produced in the $S$-wave. 

 Integration over both the solid angles of the $K^-$ and the photon leads to the following expression for the effective mass differential cross section:
\begin{equation}
\label{dsigdm}
\frac{d\sigma}{dm}= \frac{m(s-m^2)v}{24(2\pi)^3\sqrt{s(s-4 m_e^2)}}~ \left(1+\frac{2 m_e^2}{s}\right)~U,
\end{equation} 
where $v=\sqrt{1-4m_{\rm K}^2/m^2}$ is the kaon velocity in the 
$K^+K^-$ center-of-mass frame and 
\begin{eqnarray} 
\label{U}
\nonumber
U=\left(\frac{e^3}{s}\right)^2 |F_K(s)|^2~|T_{K^+ K^-}(m)|^2
 |I(m)-I(m_{\phi})|^2.
\end{eqnarray}
This effective mass distribution depends on the modulus of the $K^+K^-$
amplitude so it is not sensitive to its phase. However, the phase of the $K^+K^-$ amplitude is experimentally accessible in studies of its interference with the initial- or final-state photon radiation amplitudes.
\begin{figure}
\centering
\includegraphics[height=5.5cm]{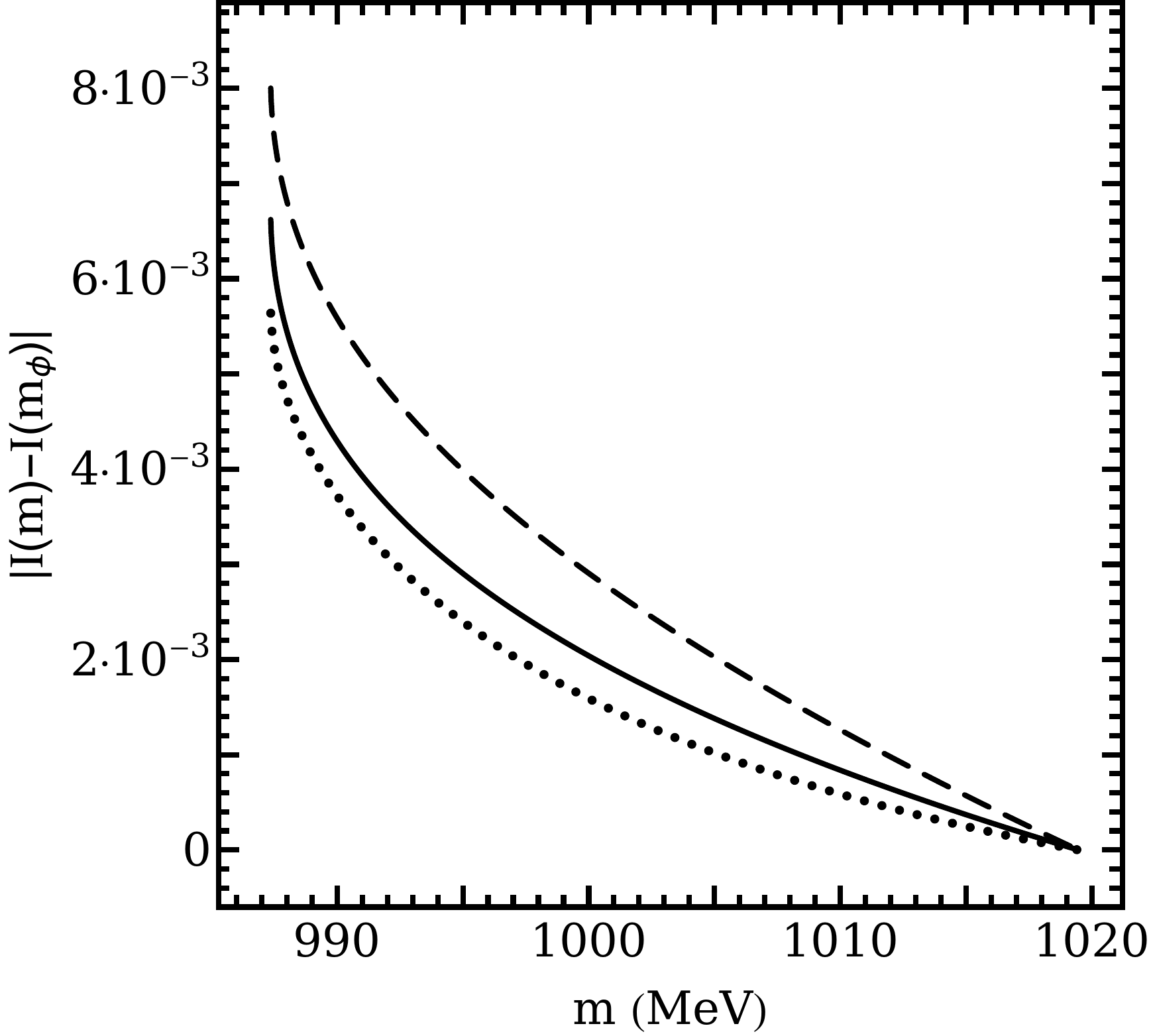}
\includegraphics[height=5.5cm]{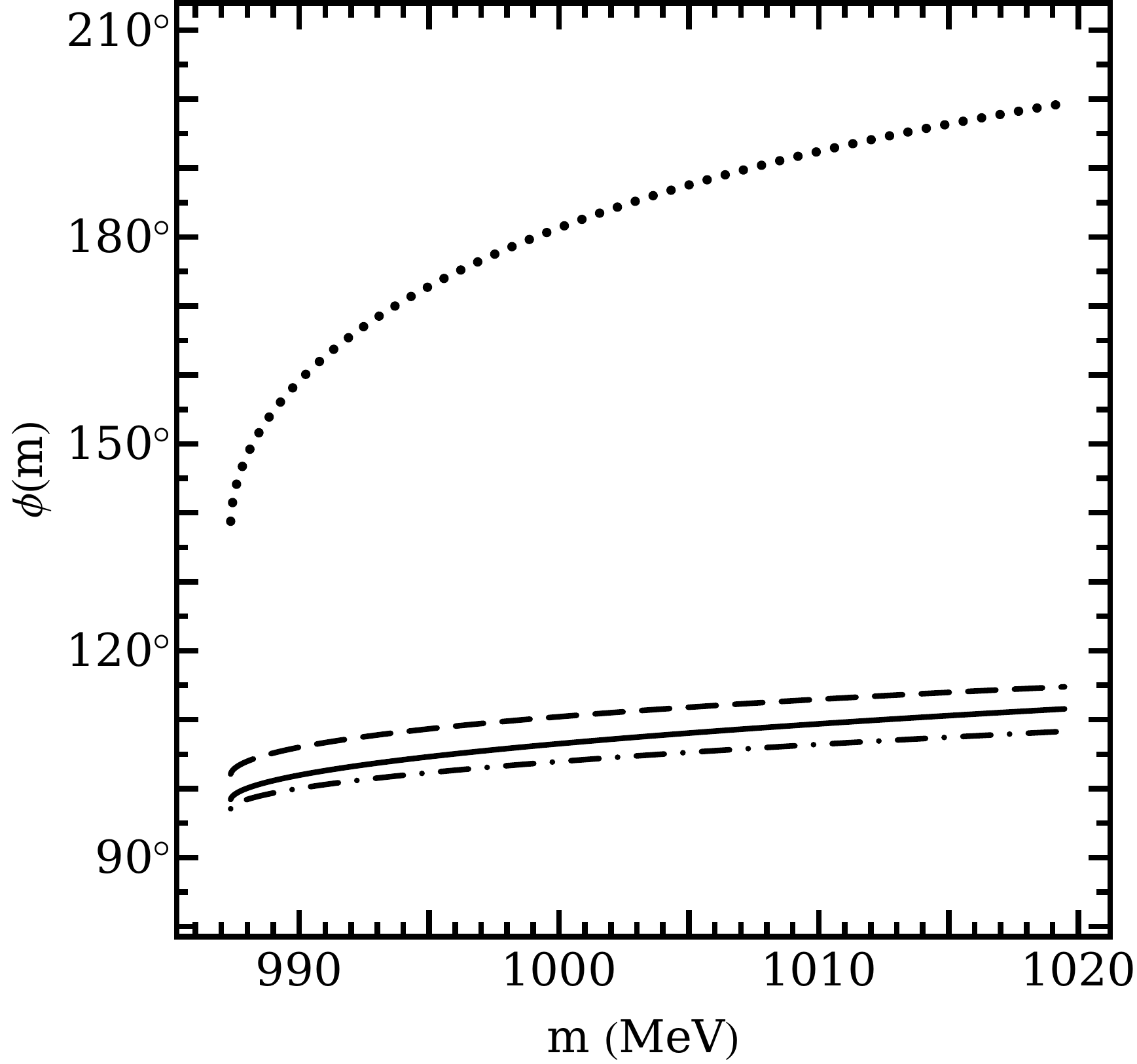}
\caption{$K^+K^-$ effective mass dependence of the modulus of the kaon-loop function $I(m_{\phi})-I(m)$ (upper panel) and its phase $\phi(m)$ (lower panel). The solid line corresponds to the function $g(k)$ (Eq.~\ref{gYama}) with the parameter $\beta \approx 1.5$ GeV and the cut-off $k_{\rm{cut}}=1$~GeV (case 1), the dotted line - to $g(k)\equiv \phi(k)$ given by Eq.~(\ref{fik}) (case 2) and the dashed curve - to $g(k)$ from Eq.~(\ref{giek}) (case 3). The dashed-dotted line in the lower panel shows the phase of the function $g_R(m)$ from Ref.~\cite{Ivan}.
\label{fig-3}}
\end{figure} 
At $s$ values close to 1 GeV$^2$ the kaon electromagnetic form factor is strongly dominated by the $\phi(1020)$ meson contribution. Then the differential cross-section for the $e^+ e^- \to K^+ K^- \gamma $ reaction can be related to the differential 
branching fraction for the $\phi(1020) \to K^+ K^- \gamma $ decay. 
The relevant square of the matrix element summed over the photon helicities and averaged over the $\phi(1020)$ helicities reads:
\begin{eqnarray}
\label{Mfi}
\nonumber
|\mathcal{M}(\phi \to K^+ K^- \gamma)|^2= e^2 g^2_{\phi K^+K^-}
|T_{K^+ K^-}(m)|^2\\
\times |I(m)-I(m_{\phi})|^2 \frac{1}{2}s(1+\cos^2\theta_{\gamma}),
\end{eqnarray}
where $g^2_{\phi K^+K^-}$ is the $\phi$ meson coupling constant to $K^+K^-$. This expression is valid if one uses the same set of diagrams
as shown in Fig.~\ref{fig-2} with a replacement of the virtual photon
$\gamma^*$ by the $\phi$ meson in the initial state. 
The differential branching fraction for the $\phi(1020) \to K^+ K^- \gamma$ decay is proportional to $|\mathcal{M}(\phi \to K^+ K^- \gamma)|^2$ as follows:
\begin{eqnarray}
\label{Br}
\begin{gathered}
d Br(\phi \to K^+ K^- \gamma) =\\
\frac{(2 \pi)^4}{2 m_{\phi} \Gamma_{\phi}}|\mathcal{M}(\phi \to K^+ K^- \gamma)|^2 d \Phi_3,
\end{gathered}
\end{eqnarray}
where $\Gamma_{\phi}$ is the total $\phi$ width. 
Inspection into Eqs.~(\ref{dsig}, \ref{sumee}, \ref{Mfi}) and (\ref{Br})
leads to the following relation between the cross-section at $s \approx m_{\phi}^2$ and the branching fraction:
\begin{eqnarray} 
\label{bratio}
\nonumber
\sigma(e^+ e^- \to K^+ K^- \gamma,~ s \approx m_{\phi}^2) \approx \sigma (e^+e^- \to \phi)\\
\times Br(\phi \to K^+ K^- \gamma),
\end{eqnarray}
where the total cross-section for the transition $e^+e^- \to \phi$, averaged over the electron and positron helicities
and summed over the $\phi$ meson spin projections, is given by
\begin{equation}
\label{eefi}
\sigma (e^+e^- \to \phi)=\frac{\Gamma_{\phi} e^4 |F_K(m_{\phi}^2)|^2}{m_{\phi}^3 g^2_{\phi K^+K^-}}.
\end{equation} 
The formula (\ref{bratio}) is valid for the differential as well as for the total cross-sections or the branching fractions. Here we consider a case of unpolarized $e^+e^-$ beams.

\section{Numerical results for the reaction $e^+ e^- \to K^+ K^- \gamma$}
\label{numK+K-}

The differential cross section for the reaction $e^+ e^- \to K^+ K^- \gamma$ 
(Eq.~\ref{dsig}) depends on the matrix element squared which in turn is proportional to the form of the loop function $I(m)-I(m_{\phi})$ (Eq.~\ref{suma}).

The modulus and the phase $\phi(m)$ of $I(m_{\phi})-I(m)$
for the four different choices of the function $g(k)$ are shown in Fig.~\ref{fig-3}. 
One observes some sensitivity of
$|I(m)-I(m_{\phi})|$ to the form of the function $g(k)$.
Although we see some difference between the phases $\phi(m)$ shown by the solid and dashed-dotted lines on the lower panel of Fig.~\ref{fig-3}, the line showing the modulus of the function $g_R(m)$ from Ref.~\cite{Ivan} after a proper rescaling it to
the form of $|I(m)-I(m_{\phi})|$ is practically indistinguishable from the solid line in the upper panel.
We have also calculated the kaon-loop function $I_{\rm r}(m)$ given by Eq.~(\ref{Ir}) using the function $g(k)$ from Eq.~(\ref{gYama}) with the parameter $\beta\approx 1.5$ GeV. The corresponding curves are very close to those given by solid lines in Fig.~\ref{fig-3}, the relative differences do not exceed 1.5 \%.

From the lower panel of Fig.~\ref{fig-3} one can see that the three curves are rather close to each other showing a dominance of the modulus of the imaginary part of $I(m_{\phi})-I(m)$ over the corresponding real part. If the real part would be zero then the phase would be 90$^0$.
There is one exception, namely the dotted curve shows a dominance of the real part over the imaginary part. This curve corresponds to the function $g(k)\equiv \phi(k)$ taken from Eq.~(\ref{fik}), normalized to 1 at $k=0$.
As we have explained in Subsection \ref{otherappr}, this function should be normalized to 1 at the running kaon momentum $k_f$, and not at $k=0$,
which is a case valid only at the $K^+K^-$ threshold. After a proper normalization of $g(k)$ in Eq.~(\ref{giek}) one obtains the dashed curve with much smaler real part of $I(m_{\phi})-I(m)$.

As we see in Eq.~(\ref{suma}), the matrix element squared defined in Eq.~(\ref{av}) is proportional to the square of the modulus of the kaon form factor $F_K(s)$. In the calculations presented below we use its parameterization by Bruch, Khodjamirian and Kuhn (input values of parameters are written in Table 2 of Ref.~\cite{Bruch} for the constrained fit). At $s=m_{\phi}^2$ one gets $|F_K(s)|^2=6287$.

\begin{figure}
\centering
\includegraphics[width=6cm]{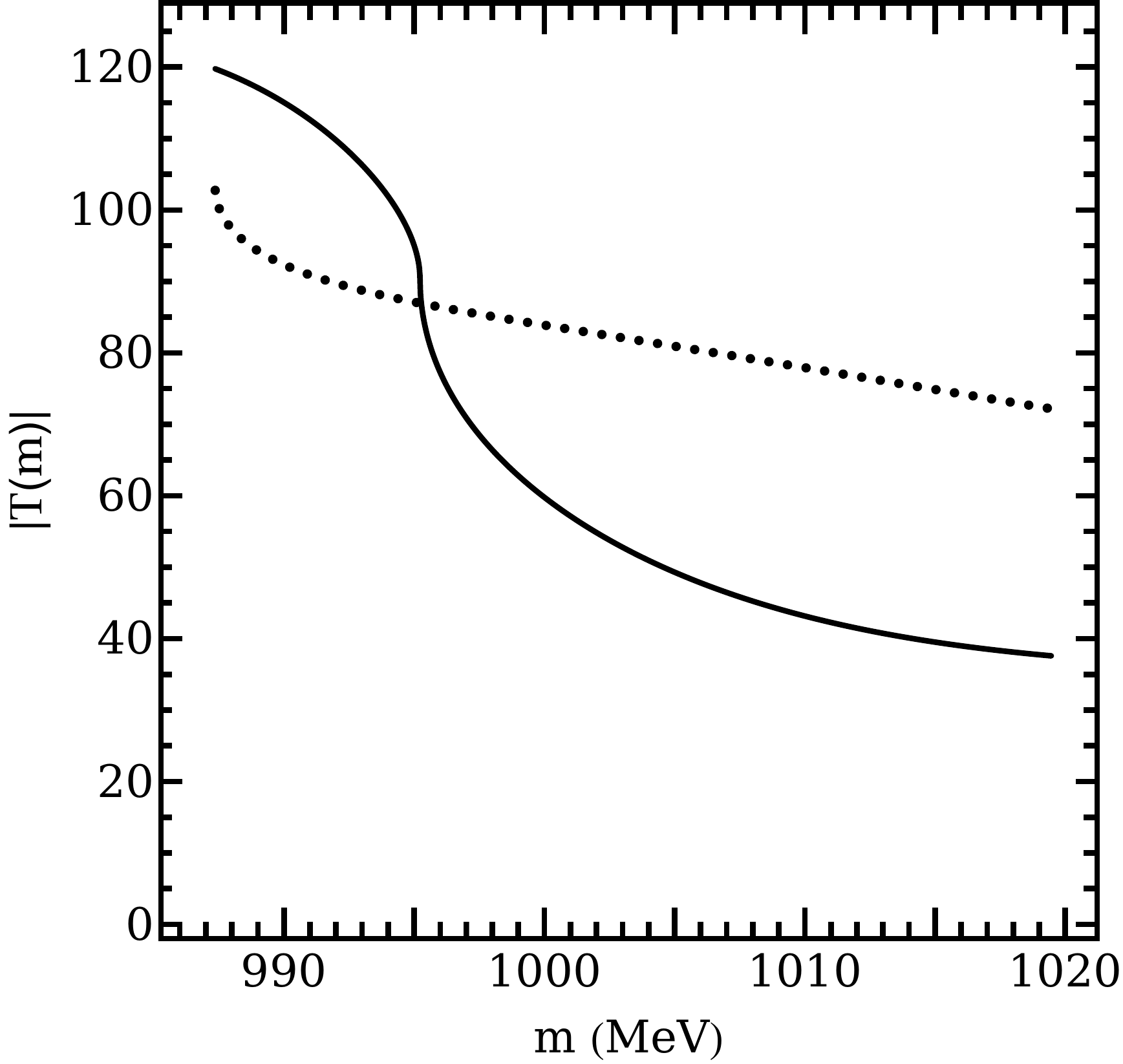}
\includegraphics[width=6cm]{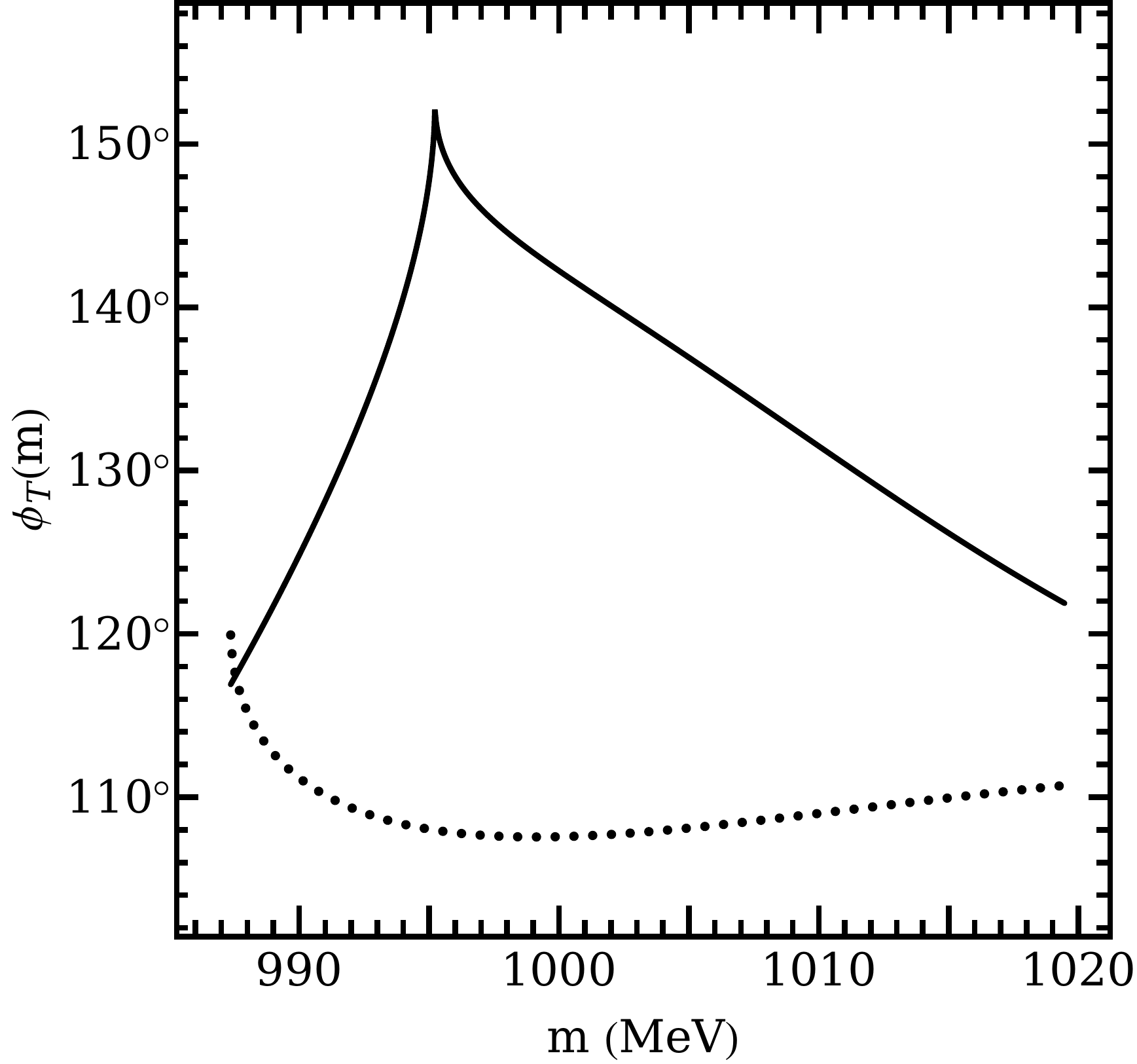}
\caption{Effective mass dependence of the moduli of the $K \bar{K}$ scattering amplitudes 
$|T(m)|$ (upper panel) and theirs phases $\phi_T(m)$ (lower panel). 
The dotted lines correspond to the elastic scattering $K^+K^- \to K^+K^-$ and the solid lines to the transition amplitude $K^+K^- \to K^0 \bar{K^0}$.}
\label{fig-4}       
\end{figure}

   The modulus and the phase of the $K^+K^-$ elastic scattering amplitudes are plotted in Fig.~\ref{fig-4} as dotted lines.
One observes somewhat steeper behaviour of these functions  near the $K^+K^-$ threshold situated at $m\approx 987.4$~MeV. This is a direct influence of the $f_0(980)$ resonance located in vicinity of the threshold.    
The solid lines drawn for the transition amplitude $T_{K^+K^- \to K^0 \bar{K^0}}$ are described in Sec.~\ref{DK0K0} where the numerical results for the reaction $e^+ e^- \to K^0 \bar{K^0} \gamma$ are presented.

\begin{figure}
\includegraphics[width=5.86cm]{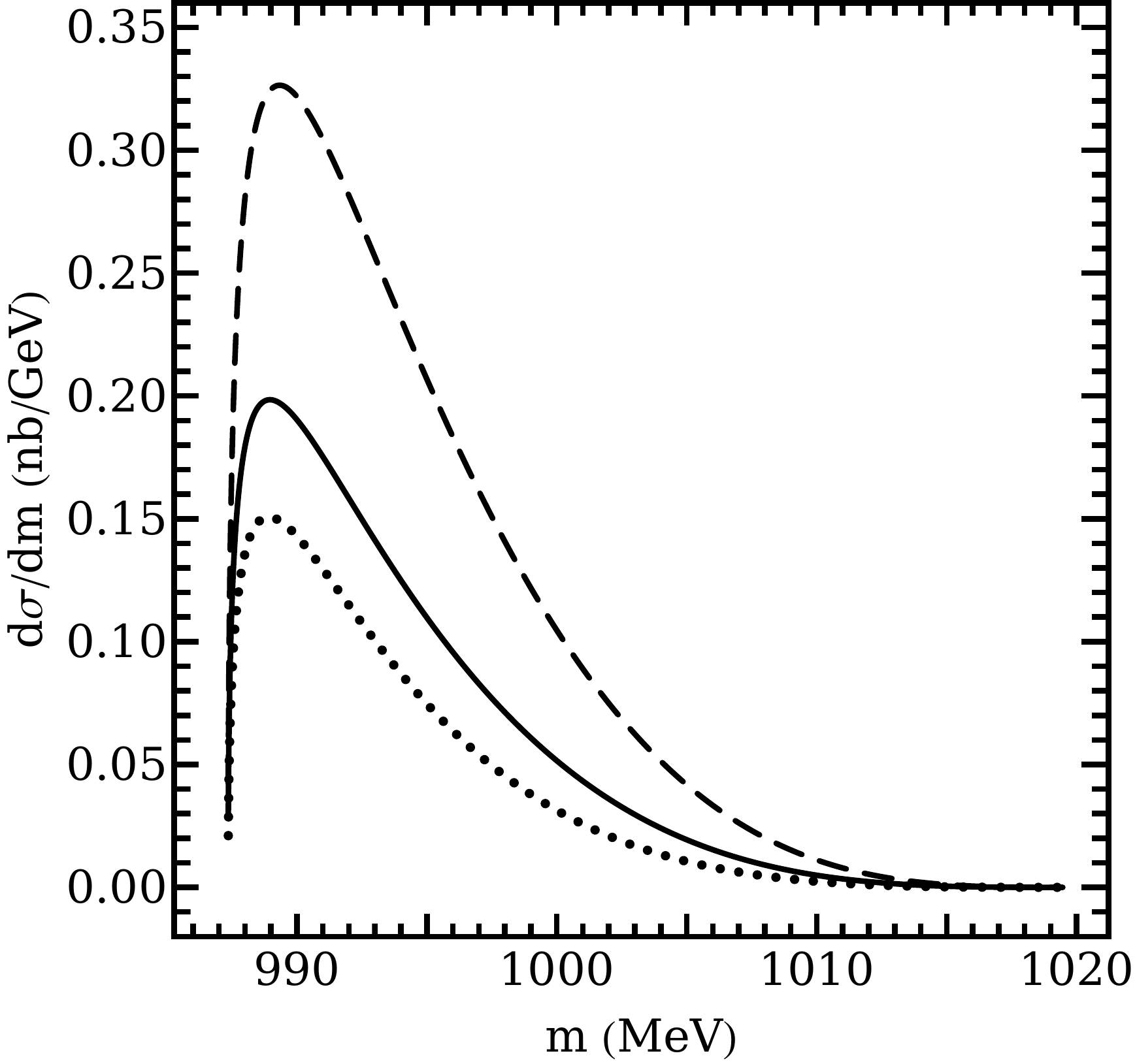}
\includegraphics[width=5.86cm]{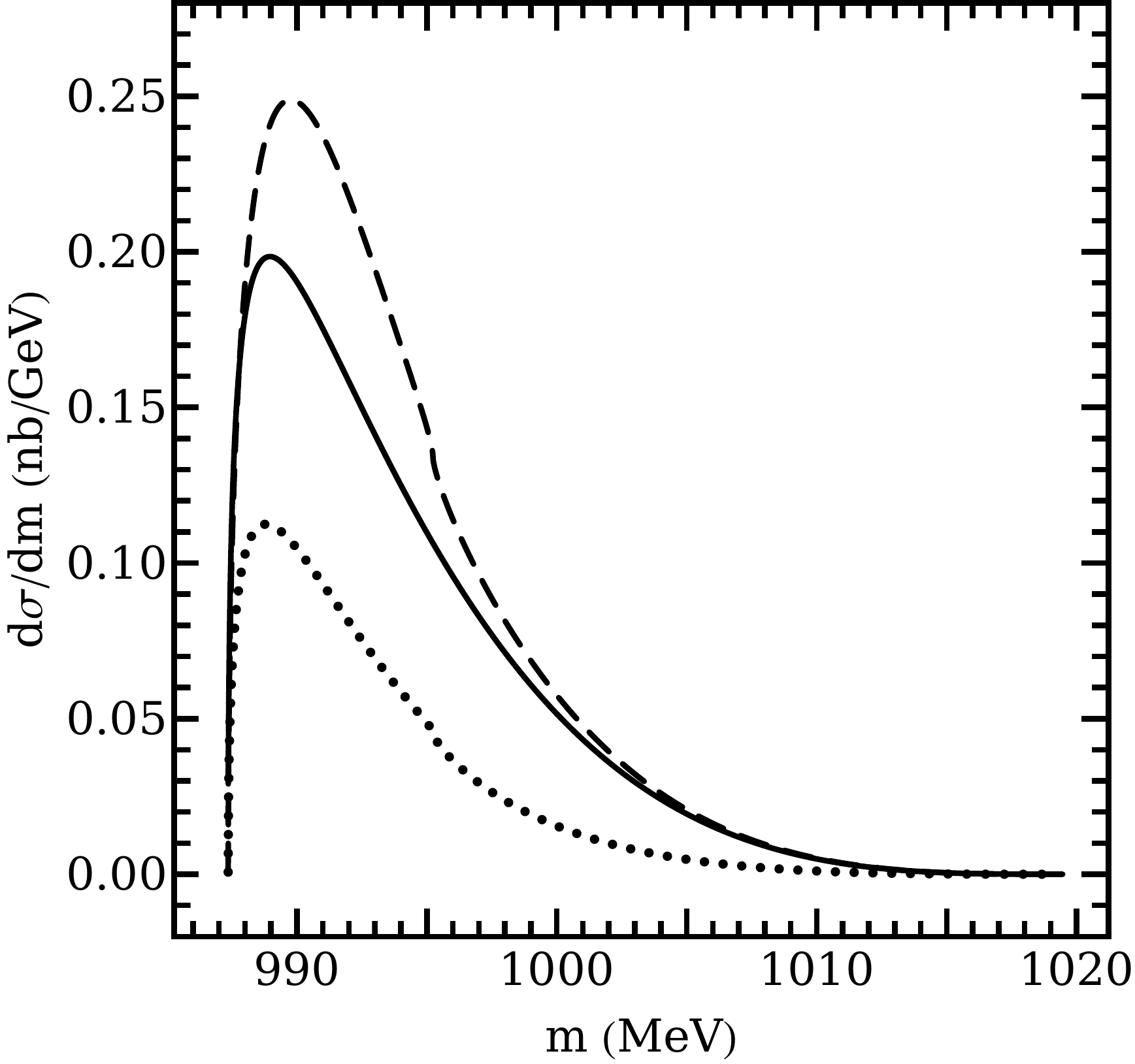}
\caption{Differential cross-section for the reaction $e^+ e^- \to K^+ K^- \gamma$ at $\sqrt{s}=m_{\phi}$ as a function of the $K^+K^-$ effective mass $m$.
 Upper panel: the lines are labelled as in the upper panel of Fig.~\ref{fig-3}
 (cases 1 to 3);
lower panel: the dashed line is calculated for the no-structure model (Ref.~\cite{NS}) (case 4), the dotted one for the kaon-loop model with parameters obtained in Ref.~\cite{KLOE} (case 5) and the solid line is the same as in the upper panel but with a different vertical scale.}
\label{fig-5}       
\end{figure}

The $K^+K^-$ effective mass distributions at $\sqrt{s}=m_{\phi}$ are plotted in Fig.~\ref{fig-5}.
One can notice that the mass distributions depend on the function $g(k)$ which influences the loop function difference
$I(m)-I(m_{\phi})$. 
All the curves have a maximum near 990 MeV, only a few MeV above the $K^+ K^-$ threshold. In the upper panel its value varies between 0.15 and 0.33 nb/GeV. Here we plot three curves calculated using the model derived in this article. 
For the function $I_{\rm r}(m)$ defined in Eq.~(\ref{Ir}), using the separable potentials to calculate the $K^+K^-$ amplitudes as described in Sec.~\ref{KKamplitudes}, the resulting $K^+K^-$ effective mass distribution 
differs relatively by 0.8 \% to 3.6 \% from the distribution shown as solid line which corresponds to the loop function $I(m)$ from Eq.~(\ref{I1}).
Therefore the corresponding curve is not plotted since it would overlap with the solid line. 

In the lower panel of Fig.~\ref{fig-5} one can see a comparison of the effective mass distributions corresponding to three different models. The dashed line has been  calculated by us for the parameters of the no-structure model (Ref.~\cite{NS}),
read from Table 1 of Ref.~\cite{KLOE}, where the analysis of the data on the $\phi \to f_0(980) \gamma \to \pi^+ \pi^-\gamma$ has been performed. Similarly, the dotted line 
has been calculated for the parameters of the kaon-loop model of Ref.~\cite{mod3} fitted in the same KLOE analysis. The solid line
is our result copied from the upper panel in order to make a more direct comparison of the results. We see that the shape of the distributions is quite similar and the maximum value of the cross section changes between about 0.11 and 0.25 nb/GeV.
Unfortunately no experimental data on the branching fraction of
$\phi \to K^+ K^- \gamma$ exist so one cannot make a direct comparison of the model results with data.
\begin{table}
\label{tab1nowy}
\caption{Values of the total cross section $\sigma_\mathrm{tot}$ and the branching fraction $Br$
for the decay $\phi \to K^+ K^- \gamma$.}
\begin{center}
\begin{tabular}{ l  c  c  c  c}
\hline			
Case &$\sigma_\mathrm{tot}$ (pb)&$Br$&Remarks\\
\hline
1&$1.85$&$4.47\cdot 10^{-7}$&loop function from Eq.~(\ref{I1})\\
& & & with $k_{\rm{cut}}=1$~GeV\\
1 r&$1.82$&$4.39\cdot 10^{-7}$&loop function from Eq.~(\ref{Ir})\\
2&$1.29$&$3.10\cdot 10^{-7}$&g(k) from Eq.~(\ref{fik})\\
3&$3.37$&$8.13\cdot 10^{-7}$&g(k) from Eq.~(\ref{giek})\\
4&$2.29$&$5.51\cdot 10^{-7}$&no-structure model~\cite{NS},~\cite{KLOE}\\ 
5&$0.85$&$2.05\cdot 10^{-7}$&kaon-loop model~\cite{mod3},~\cite{KLOE}\\
 \hline  
\end{tabular}
\end{center}
\end{table}


Calculation of the total reaction cross-section for the $e^+ e^- \to K^+ K^- \gamma$ transition, by integration of the differential cross-section $d\sigma/dm$ over the $m$-range
from the $K^+K^-$ threshold up to $m_{\phi}$ leads to the values shown in Table~I. 
In the same table we give the corresponding branching fractions for the $\phi$ meson decay into $K^+ K^- \gamma$.
The values in the five rows correspond to the five cases defined
in the captions of Figs. \ref{fig-3} and \ref{fig-5}.
In the row labelled by $1~\rm r$ we show the values calculated for the kaon-loop function $I_{\rm r}(m)$ from Eq.~(\ref{Ir}). They differ by less than 2 \% from the corresponding values shown in the first row (case 1). 

In calculations of the branching fractions
we have used the value of 4.15 $\mu$b for the total 
$e^+ e^- \to \phi$ cross-section at the $\phi$ resonance peak position (Eq.~\ref{eefi}). 
Early calculations of the branching fraction for the reaction $\phi \to \gamma (a_0+f_0) \to \gamma K^+ K^-$ have given the values between
$2.0 \cdot 10^{-7}$ and $2.6 \cdot 10^{-6}$ ~\cite{Ivan}.
In Ref.~\cite{KL} the $\phi$ radiative decays into $K^+K^-$ have been examined using the function $\phi(k)$ (Eq.~\ref{fik}) from Ref.~\cite{Close} with an estimate of the branching fraction $Br(\phi  \to \gamma(f_0+a_0) \to \gamma K^+K^-) \le 10^{-6}$.
In Ref.~\cite{Aczasow2001} one finds the values $2.25 \cdot 10^{-6}$ 
and $8.12 \cdot 10^{-7}$ in two model variants of the $f_0$ and $a_0$
positions and coupling constants.
 All the values given in Table I are below $10^{-6}$.
The dispersion of the values comes from two factors: the type of the
kaon-loop function $I(m)$ and the form of the $K^+K^-$ scattering amplitude
(see Eq.~(\ref{Mfi}) for the reaction matrix element).
\section{Reactions with other meson pairs in the final states}
\label{other}
There is a natural way to generalize the amplitudes derived for the 
$e^+ e^- \to K^+ K^- \gamma$ reaction to other reactions like
$e^+ e^- \to P_1 P_2 \gamma$ where in the final state pairs of the
pseudoscalar mesons $P_1$ and $P_2$ are produced. In the intermediate state
the same $K^+ K^-$ loop is present. In order to write down the amplitudes corresponding to the $e^+ e^- \to P_1 P_2 \gamma$ reaction and to follow the derivation presented above for the $e^+ e^- \to K^+ K^- \gamma$ reaction, one has to replace in each step the elastic $K^+ K^-$ amplitude
$T(k)$ in Eq.~(\ref{Tk}) by the inelastic $K^+ K^-\to P_1 P_2 $ amplitude
\begin{eqnarray}
\begin{gathered} 
\label{TP1P2}
T_{K^+ K^-\to P_1P_2}(k)=\\
\langle P_1(k_1) P_2(k_2)|~\tilde{T}(m)~|~K^-(-k+p-q)~ K^+(k)~\rangle .
\end{gathered}
\end{eqnarray}
There are several $P_1P_2$ states coupled to the $K^+K^-$ channel.
Below one can enumerate some of them:\\
1. $e^+ e^- \rightarrow \pi^+ \pi^- \gamma$,\\
2. $e^+ e^- \rightarrow \pi^0 \pi^0 \gamma$,\\
3. $e^+ e^- \rightarrow \pi^0 \eta \gamma$,\\
4. $e^+ e^- \rightarrow K^0 \bar{K^0} \gamma$,\\
5. $e^+ e^- \rightarrow K^+ K^- \gamma$.\\
All these channels can be simultaneously studied in a unitary way when the operator $\tilde{T}(m)$ becomes a reaction matrix describing all the possible transitions between the $P_1~P_2$ pairs of mesons.
The on-shell amplitude equivalent to that written in Eq.~(\ref{norm}) is given by
\begin{eqnarray} 
\label{P1P2norm}
\begin{gathered}
T_{K^+K^-\to P_1P_2}=\\
\frac{4\pi m}{i \sqrt{k_f k_{12}}}(S_{_{K^+ K^-\to ~ P_1 P_2}}-
\delta_{K^+K^-,~ P_1 P_2}),
\end{gathered}
\end{eqnarray}
where we have introduced the $S$-matrix element corresponding to the reaction 
$K^+ K^-\to~ P_1 P_2$. In the above equation $k_{12}$ denotes the relative momentum of the $P_1$ and $P_2$ particles in their center-of-mass system.
Then, extending the model constructed for a description of the reaction $e^+ e^- \to K^+ K^- \gamma$ to other reactions, one can perform a coupled channel analysis of the whole set of the reactions $e^+ e^- \to \phi(1020) \to P_1 P_2 \gamma$. 

\begin{table}
\label{tab2nowy}
\caption{Values  of the total cross section $\sigma_\mathrm{tot}$ and the branching fraction $Br$
for the reaction $e^+ e^- \to K^0 \bar{K^0} \gamma$.}
\begin{center}
\begin{tabular}{ l  c  c c}
  \hline			
Case&$\sigma_\mathrm{tot}$ (pb)&$Br$&  Remarks\\
\hline
1&$1.67 \cdot 10^{-1}$&$4.03 \cdot 10^{-8}$&loop function from Eq.~(\ref{I1})\\
& & & with $k_{\rm{cut}}=1$~GeV\\
1 r&$1.65 \cdot 10^{-1}$&$3.98 \cdot 10^{-8}$&loop function from Eq.~(\ref{Ir})\\
2&$1.02 \cdot 10^{-1}$&$2.46 \cdot 10^{-8}$&g(k) from Eq.~(\ref{fik})\\
3&$3.38 \cdot 10^{-1}$&$8.16 \cdot 10^{-8}$&g(k) from Eq.~(\ref{giek})\\
  \hline  
\end{tabular}
\end{center}
\end{table}
\section{Description of the reaction $e^+ e^- \to K^0 \bar{K^0} \gamma$}
\label{DK0K0}
 As the first step in a derivation of the amplitude for the reaction
$e^+ e^- \to K^0 \bar{K^0} \gamma$, we make the following isospin decomposition of the $K^0 \bar{K^0}$ state:
\begin{equation}
\label{K0K0}
|K^0 \bar{K^0}\rangle=\frac{1}{\sqrt{2}}(|I=0\rangle -~|I=1\rangle).
\end{equation}
The on-shell transition amplitude from the $K^+K^-$ to the  
$K^0 \bar{K^0}$ state can be expressed as 
\begin{equation}
\label{tra}
T_{K^+K^- \to~ K^0 \bar{K^0}}~(m)=\frac{1}{2}[t_0(m)-t_1(m)],
\end{equation}
where the amplitudes $t_0(m)$ and $t_1(m)$ have been introduced in Eq.~(\ref{TI}).
Like in Sec.~\ref{KKamplitudes} for the elastic $K^+K^-$ amplitudes, in the numerical calculations of the amplitude $t_0(m)$ in Eq.~(\ref{tra})
we have made a shift of the mass $m$ by about 2 MeV by changing the argument $m$ into $m~ (m_{\rm {av}}/m_{K^0})$. 
The separable potential parameters of the amplitude $t_1(m)$ have been directly adjusted to the value of the neutral kaon mass.

The amplitude for the reaction $e^+ e^- \to K^0 \bar{K^0} \gamma$ can be obtained from Eq.~(\ref{Atot2}) after the substitution of $T_{K^+K^- \to~ K^0 \bar{K^0}}~(m)$ in place of $T_{K^+ K^-}(m)$.

The next steps needed in calculation of the cross section for the reaction $e^+ e^- \to K^0 \bar{K^0} \gamma$ are exactly the same as given in Sections \ref{model} and \ref{Diff} for the $e^+ e^- \to K^+ K^- \gamma$ process but
again the substitution of $T_{K^+ K^-}(m)$ by $T_{K^+K^- \to~ K^0 
\bar{K^0}}~(m)$ has to be done in Eqs.~(\ref{suma}),~(\ref{sumee}) and (\ref{U}).
As a result of these replacements Eq.~(\ref{dsigdm}) gives the differential cross section for the reaction $e^+ e^- \to K^0 \bar{K^0} \gamma$ if we also change $v$ into $v_0=\sqrt{1-4m_{K^0}^2/m^2}$, the $K^0$ velocity in the $K^0 \bar{K^0}$ center-of-mass frame.
\begin{figure}[ht]
\includegraphics[width=5.86cm]{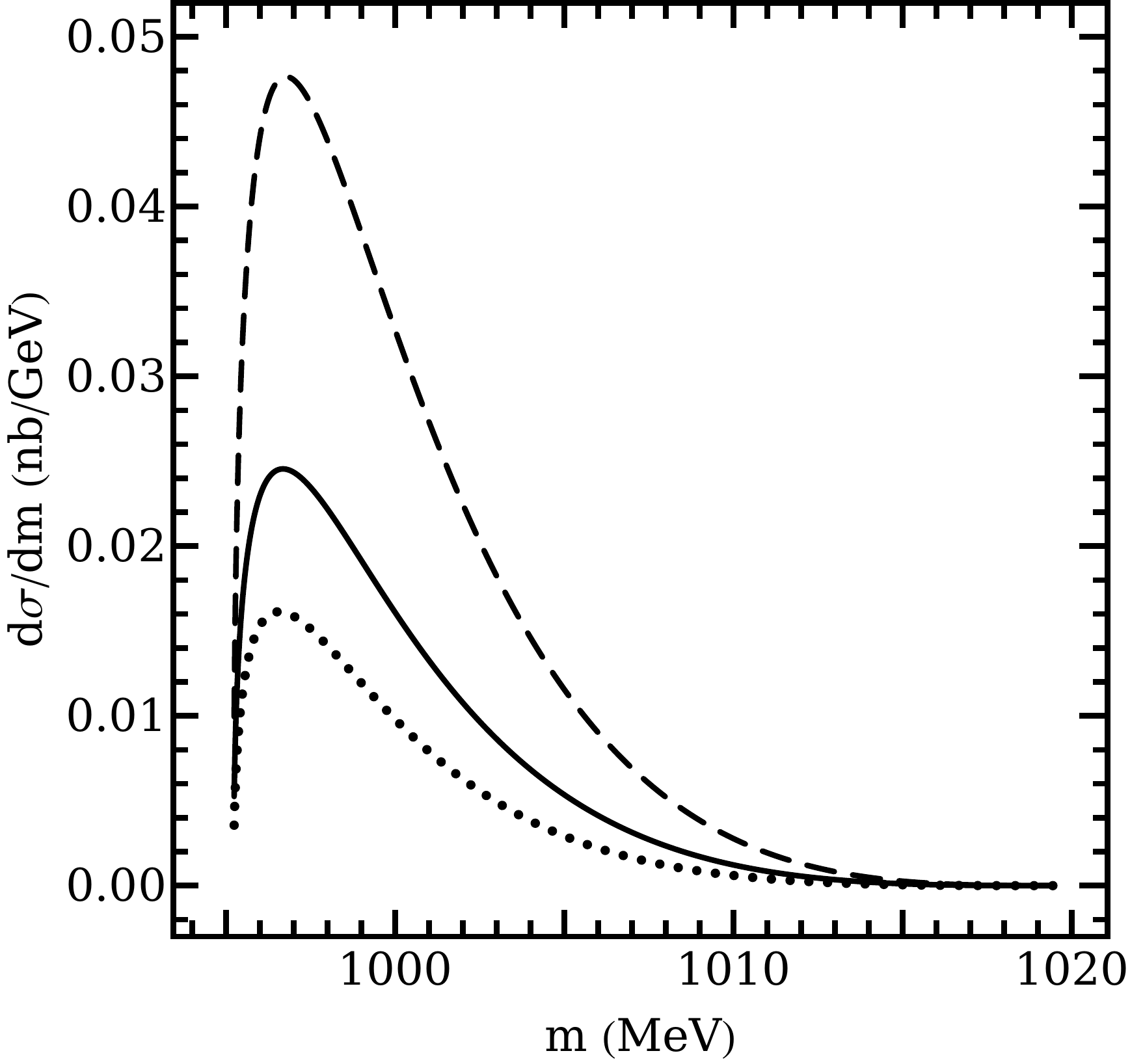}~~

\caption{Differential cross-section for the reaction $e^+ e^- \to K^0 \bar{K^0} \gamma$ at $\sqrt{s}=m_{\phi}$ as a function of the $K \bar{K}$ effective mass.
The lines are labelled as in the upper panel of Fig.~\ref{fig-5}
(cases 1 to 3).}
\label{fig-6}       
\end{figure} 

The $K^0 \bar{K^0}$ effective mass differential cross sections are shown in Fig.~\ref{fig-6}.
One observes a considerable lowering of the cross section values in comparison with Fig.~\ref{fig-5}.
This fact has two reasons. 
The first one is related to the limited phase space for the 
$e^+ e^- \to K^0 \bar{K^0} \gamma$ reaction in comparison with the phase space of the  $e^+ e^- \to K^+ K^- \gamma$ reaction.
Simply speaking, this is due to the value of the $K^0 \bar{K^0}$ threshold mass (about $995.2$ MeV) which is by 7.8~MeV higher than the $K^+ K^-$ threshold mass.
The second reason is illustrated in the upper panel of Fig.~\ref{fig-4}
where we see that at the effective mass larger than the $K^0 \bar{K^0}$ threshold the modulus of the amplitude $T_{K^+K^-\to~K^0 
\bar{K^0}}~(m)$ is substantially lower than the modulus of the elastic $K^+K^-$ amplitude.
As seen in the lower panel of Fig.~\ref{fig-4}, the phases of both amplitudes are also different.
However, this difference does not generate a further effect
on the values of the differential cross sections which are proportional to  the square of the amplitude moduli (see, for example, Eqs.~(\ref{dsigdm}) and (\ref{U})).
Here one can mention that a characteristic "horn" shape of the solid line is due to the opening of the $K^0 \bar{K^0}$ threshold near $m=995$ MeV. We have
calculated the phase below 995 MeV by making an analytical continuation of the transition amplitude for the reaction
$K^+K^- \to~ K^0 \bar{K^0}$ beyond its physical threshold. 
One has to add here that the differential cross section for the kaon-loop function $I_{\rm r}(m)$ (Eq.~\ref{Ir}) is very similar to that shown in Fig.~\ref{fig-6} by solid line. The relative differences vary between 0.9 \% and 2.4 \%, so once again we do not show the corresponding line in order to evict almost a complete overlap of lines.

The calculated results for the total reaction cross-sections of the $e^+ e^- \to K^0 \bar{K^0} \gamma$ transition and the branching fractions for the $\phi$ decay into 
$K^0 \bar{K^0} \gamma$ are given in Table II.
\begin{figure}[ht]
\includegraphics[width=5.86cm]{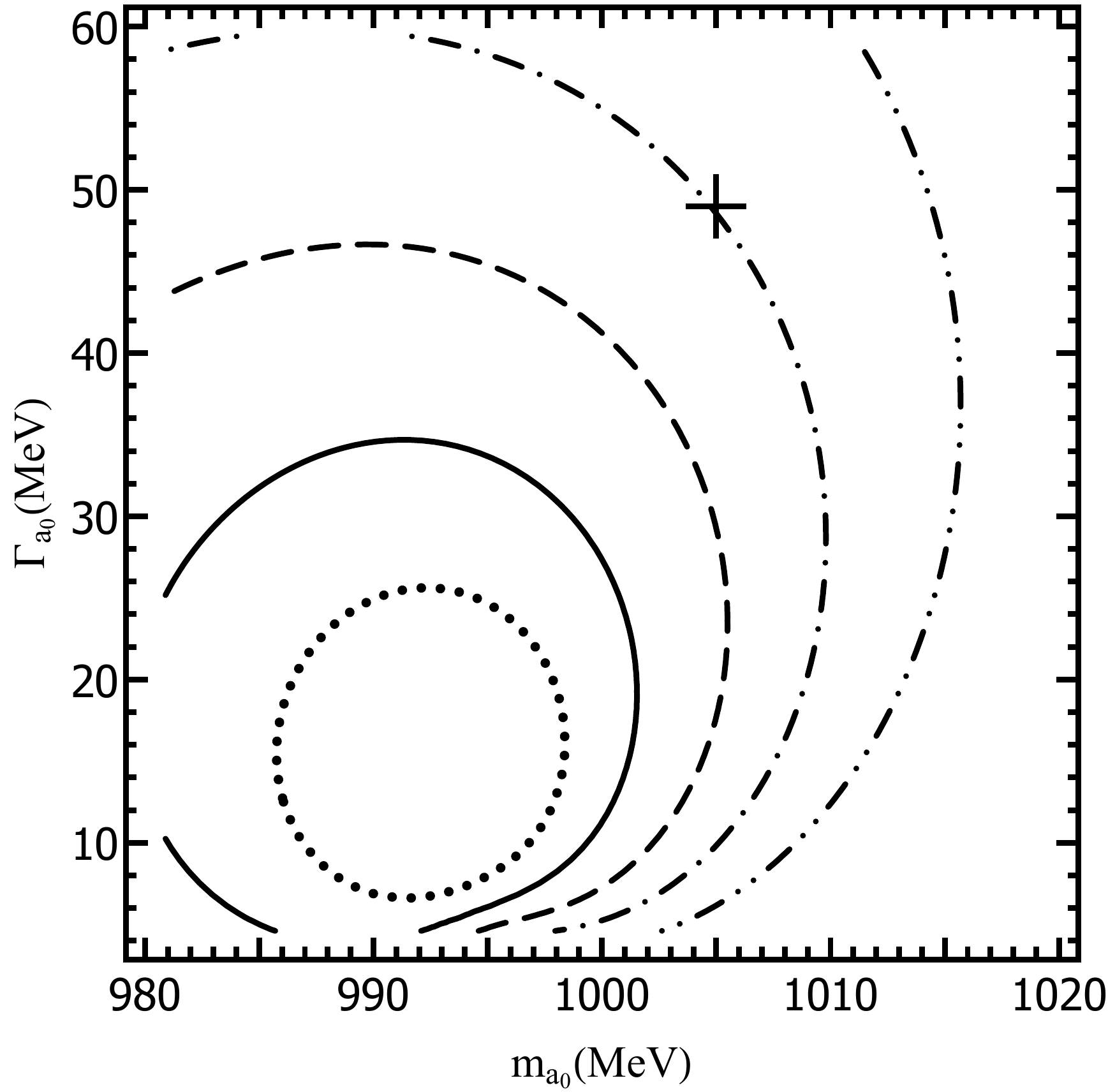}~~

\caption{Contours of the branching fraction $Br$ for the decay $\phi \to K^0 \bar{K^0} \gamma$ in the complex plane of the $a_0(980)$ pole position:
$m_{a_0(980)}$ is the resonance mass and $\Gamma_ {a_0(980)}$ its width.
The solid curve corresponds to the KLOE upper limit $Br=1.9 \cdot 10^{-8}$,
the dotted one to $Br=1.0 \cdot 10^{-8}$, the dashed curve to 
$Br=3.0 \cdot 10^{-8}$, the dashed-dotted one to $Br=4.0 \cdot 10^{-8}$ , and the dashed- double dotted one to
$Br=5.0 \cdot 10^{-8}$. The cross indicates the $a_0(980)$ resonance position
on sheet $(-+)$ found in Ref.~\cite{Furman}.
}
\label{fig-7}       
\end{figure}  
By comparison with Table I one sees that the branching fractions for the $\phi$ decays into $K^0 \bar{K^0} \gamma$ are by about one order of magnitude smaller than the corresponding branching fractions for the transition $\phi \to K^+K^- \gamma$.
The four cases shown in Table II correspond to the same cases seen in Table I. Once again, we notice a small difference between the case 1 and the case $1~\rm r$. The relative difference is only about 1.2 \%.

The results presented in Table II can be compared with the values for the branching fraction of the $\phi$ decay into the $K^0 \bar{K^0} \gamma$ channel calculated using different models. 
Here we quote some of them: in Ref.~\cite{Ivan} one finds the values $2.0 \cdot 10^{-9}$ and $1.3 \cdot 10^{-8}$, Bramon, Grau and Pancheri have obtained $7.6 \cdot 10^{-9}$ in~\cite{Bramon1}, the results of Achasov and Gubin from Ref.~\cite{Aczasow2001} are $4.36 \cdot 10^{-8}$
and $1.29 \cdot 10^{-8}$, Oller in~\cite{Oller2} gave the values $3.7\cdot 10^{-8}$ or $6.43 \cdot 10^{-9}$ depending on the $K \bar{K}$
amplitude type, the result of Escribano from Ref.~\cite{Escribano}
is $7.5\cdot 10^{-8}$. 

The upper limit $1.9 \cdot 10^{-8}$ for the $\phi \to K^0 \bar{K^0} \gamma$ decay found by the KLOE Collaboration in Ref.~\cite{kloe2} is of the same order as the numbers in Table II.
It is possible to generate lower values of the theoretical branching fractions
 by a moderate change of the pole positions of the scalar mesons $f_0(980)$ or $a_0(980)$. 
We repeat here that for the present model of the isospin zero $K\bar{K}$ amplitude the $a_0(980)$ pole
position is given by the mass $m_{a_0(980)}=1005$ MeV and the width $\Gamma_ {a_0(980)}=49$ MeV. 
As seen in Fig.~\ref{fig-7}, calculated for the case 1 in Table II, the KLOE lower limit can be reached by changing the
position of the $a_0(980)$ resonance or its width on the sheet $(-+)$ by about 5 MeV. 
This figure indicates an important role of future experimental measurements of the reactions $e^+ e^- \to K^+ K^- \gamma$ and $e^+ e^- \to K^0 \bar{K^0} \gamma$ in a more precise determination of the properties of scalar resonances.

As a final comment we can add that the same procedure of the amplitude replacement just described for the reaction $e^+ e^- \to K^0 \bar{K^0} \gamma$ can be done for other reactions
like for the first three transitions listed below Eq.~(\ref{TP1P2}).
\section{Conclusions}
\label{Concl}
In summary, the theoretical model of the reaction amplitudes for the processes $e^+ e^- \to K^+ K^- \gamma$ and $e^+ e^- \to K^0 \bar{K^0}\gamma$ 
 has been formulated. The strong interactions between the charged and neutral kaons are included in the elastic scattering amplitude $T_{K^+ K^-}(m)$ and in the transition amplitude 
$T_{K^+K^- \to~ K^0 \bar{K^0}}~(m)$.
The formulae for the total reaction amplitude $A= A_1+A_2+A_3+A_4$, valid for the general form of the $K\bar{K}$ scattering amplitudes, are presented in Eqs.~(\ref{A1k}-\ref{A4k}) and in Eqs.~(\ref{T1}-\ref{M5}).

We have shown that some models used in past for a description of the radiative $\phi$ decays into $ K^+ K^- \gamma$ and $K^0 \bar{K^0}\gamma$ 
can be treated as special cases of the model presented in Sec.~\ref{model derivation}.
For the reaction $e^+ e^- \to K^+ K^- \gamma$, the approximate form of the amplitude, valid for small values of the outgoing photon energy,   is given by Eq.~(\ref{Atot2}).
It is proportional to $T_{K^+ K^-}(m)$ and to the kaon-loop function difference $I(m)-I(m_{\phi})$. 
The alternative form $I_{\rm r}(m)$ of the function $I(m)$ from Eq.~(\ref{I1})
can be seen in Eq.~(\ref{Ir}).
These functions depend on the off-shell
behaviour of the $K^+K^-$ elastic scattering amplitude $T_{K^+ K^-}(m)$
given by the function $g(k)$ in Eq.~(\ref{Ton}).
However, if the function $g(k)$ depends sensitively on the kaon momenta $k$ only at $k\geq 1$ GeV, then the reaction amplitude is close to the amplitude calculated in the limit of point-like kaons ($g(k)\equiv 1$).
The gauge invariance condition leading to vanishing reaction amplitude at the photon energy going to zero has an important consequence for that behaviour.

The amplitude for the reaction $e^+ e^- \to  K^0 \bar{K^0} \gamma$ can be obtained from Eq.~(\ref{Atot2}) after a replacement of $T_{K^+ K^-}(m)$ by
$T_{K^+K^- \to~ K^0 \bar{K^0}}~(m)$.
 
The formulae for the differential cross sections describing the $K\bar{K}$ effective mass distributions and the photon and kaon angular distributions have been obtained. 
The numerical calculations of the effective mass distributions, the total reaction cross sections and the branching fractions for the $\phi(1020)$ decays
into $K^+K^- \gamma$ and $K^0 \bar{K^0}$ have been performed.
The separable
meson-meson potentials with the parameters taken from Refs.~\cite{KLL} and ~\cite{Furman} have been used to calculate the $K\bar{K}$ amplitudes. 
Other forms of the kaon-kaon scattering amplitudes can be easily included in alternative studies of the same reactions.
The present model can be used in future experimental analyses of the  
reactions $e^+ e^- \to K^+ K^- \gamma$ and $e^+ e^- \to K^0 \bar{K^0} \gamma$, in particular at the $e^+ e^-$
energies close to the mass of the $\phi(1020)$ meson. We have also generalized this model to the reactions 
$e^+ e^- \to P_1 P_2 \gamma$ with pairs of the
pseudoscalar mesons $P_1$ and $P_2$ different from $K^+K^-$ or from $K^0 \bar{K^0}$.
The model in this form can serve in couple channel analyses of the $e^+ e^-$ data for the production processes  
including $K^+ K^-$, $K^0_S K^0_S$, $ \pi^+ \pi^-$, $\pi^0 \pi^0$ and $\pi^0 \eta$ pairs of mesons in the final state. Such combined analysis 
can provide a valuable information about the positions of the $a_0(980)$ and $f_0(980)$ resonances and about the near threshold $K\bar{K}$ scattering amplitudes.

At the end let us add a remark concerning a possible contribution of other intermediate states with the same quantum numbers as $K^+K^-$. The $K^+K^-$ loop dominates at the $e^+e^-$ center-of-mass energy close to the $\phi(1020)$ meson mass since the branching fraction for the $\phi$ decay into $K^+K^-$ is much larger than any other branching fraction for the decay into charged particles [1].
For example, we have checked that the $\pi^+\pi^-$ loop with the subsequent $\pi^+\pi^- \to K^+K^-$  transition leads to a reaction cross section by a factor of about $10^{-6}$ smaller than the cross section calculated with the $K^+K^-$ intermediate loop.

\section*{Acknowledgements} 
 We would like to thank Dominika Hunik, Krzysztof Kacprzak, Adam \L{}abaza, Szymon Starzonek and Tomasz Twar\'og for their participation in early stages of our studies. This work has been supported by the Polish National Science Centre (grant no 2013/11/B/ST2/04245). We are grateful to Henryk Czy\.z for a useful correspondence. 

\vspace{1cm}
\section{Appendix A: Amplitude $A_4(m)$} 
\label{Appendix A}
Let us examine a dependence of the amplitude $A_4(m)$ defined by Eqs.~(\ref{A4k3}-\ref{I4}).   
One can predict a weak dependence of the imaginary part of the function $I_4(m)$ on $\omega\equiv (m_{\phi}^2-m^2)/(2 m)$.
Essentially the integral value depends on the ratio $\omega/p_0$. In the $K^+ K^-$ center-of-mass frame the energy $p_0=\sqrt{m^2_{\phi}+\omega^2}$ and the photon energy $\omega$ has the following upper limit: 
$\omega \leq (m_{\phi}^2-4 m_{\rm K}^2)/(4 m_{\rm K})$.
This value is equal to about 32 MeV which is much smaller than the $\phi(1020)$ meson mass.
The maximum photon momentum is also smaller than the average of the kaon lower and upper momentum limit equal to $(k^{'}+k^{''})/2=p_0 v_L/2$.
This average momentum is equal to about 127 MeV and exceeds the maximum photon energy 32 MeV. The photon energy is also smaller than the typical range of the kaon momentum distribution which can be represented by the parameter $\mu$ in Eq.~(\ref{fik}), or by the range parameter $\beta$ in Eq.~(\ref{gYama}).
Let us note here that the variable $y$ present in the integrand of Eq.~(\ref{ImI4k}) 
takes the value $-k$ at $k=k^{'}$ and the value $+k$ at $k=k^{''}$, so the integrand function vanishes at both limits of $k$. 
Since the difference $k^{''}-k^{'}=\omega$, the factor $\omega$ in the denominator of Eq.~(\ref{ImI4k})
is cancelled and even in the limit $\omega \to 0$ one gets the finite value of the imaginary part of the function $I_4$ (see Eq.~(\ref{ImI4zero})).

We can also infer a weak $m$-dependence of the function $I_4(m)$ from studies of the its integrand in Eq.~(\ref{I4}).
If one makes an expansion of the sum $1/M_4(k)+1/M_5(k)$ in series of $\omega$, then after an integration over the angles of the vector
$\mathbf{k}$ the integrand depends only on even powers of $\omega$.
Since $\omega$ is small the function $I_4(m)$ varies very slowly with $m$.

The  above qualitative considerations, which indicate a weak $\omega$ dependence of 
$\mathfrak{Im}~A_4$, can be further supported by the numerical results obtained for the 
function $g(k)$ taken in the form of Eq.~(\ref{fik}) or given by Eq.~(\ref{giek}).
In the first case the relative variation of $\mathfrak{Im}~A_4$ in the whole region of $\omega$ between 0 and 32 MeV is smaller than 0.6 \%. In the latter case it is smaller than 2.1 \%.

It is also possible to estimate numerically the real part of the function $I_4(m)$ given by three-dimensional integral in Eq.~(\ref{I4}).
Like for $\mathfrak{Im}~I_4$ one observes a weak dependence on $\omega$.
If we take the function $g(k)$ given by Eq.~(\ref{gYama}) with the parameter $\beta$ of the order of 1.5 GeV, much larger than the parameter $\mu$=141 MeV used in Eq.~(\ref{fik}), then we can obtain even much weaker dependence of $A_4(m)$ than in the two cases described above.
 Therefore one can conclude that the variation of the amplitude $A_4(m)$ with $m$ is so weak that we can take for it the value at $m=m_{\phi}$:
$A_4(m)\approx A_4(m_{\phi})$.

\section{Appendix B: Real part of the kaon-loop function $I_r(m)$}
\label{Appendix B}
In this Appendix we give formulae derived for the real part of the kaon-loop function $I_{\rm r}(m)$ defined in Eq.~(\ref{Ir}):

\begin{eqnarray}
\label{ReIr}
\begin{gathered}
\mathfrak{Re}~ I_{\rm r}(m)= \frac{1}{(2 \pi)^2}~{\rm P} \int_0^{\infty} dk  ~g(k)~ k
\\ \times \left[ W_1(k)+W_2(k)+W_3(k)+W_4(k)+W_5(k)\right],
\end{gathered}
\end{eqnarray}

where
\be 
\label{W1}
W_1(k)=-\frac{4 k}{E_k(m^2-4 E_k^2)},
\ee
\be 
\label{W2}
\begin{gathered}
W_2(k)=\frac{1}{\omega E_k (m^2-4E_k^2)}\\
\times \left (2 k E_k-m_{\rm K}^2  \ln
\frac{E_k+k}{E_k-k}\right),
\end{gathered}
\ee
\be 
\label{W3}
\begin{gathered}
W_3(k)=\frac{1}{2 m \omega E_k (m-2 E_k)}\\
\times \left [2 y k + (k^2 - y^2) \ln \left |\frac{y+k}{y-k}\right|\right],
\end{gathered}
\ee
\be 
\label{W4}
\begin{gathered}
W_4(k)=-\frac{1}{2 m \omega E_k (m+2 E_k)}\\
\times\left [2 t k + (k^2 - y^2) \ln  \frac{t+k}{t-k}\right],
\end{gathered}
\ee
\begin{eqnarray}
\label{W5}
\nonumber
W_5(k)=\frac{1}{2\omega^3} [E_p-E_m+\frac{\omega^2 m_{\rm K}^2 \ln x_1 }{m E_k(m-2 E_k)} \\
\nonumber
-\frac{\omega^2 m_{\rm K}^2 \ln x_2}{m E_k(m+2 E_k)}  
+\frac{m_{\phi}^2(E_k-E')(E_k-E'')}{m E_k(m-2 E_k)} \ln x_3 \\
\nonumber
-\frac{m_{\phi}^2(E_k+E')(E_k+E'')}{m E_k(m+2 E_k)} \ln x_4 ].\\
\end{eqnarray} 
The variable $y$ present in Eq.~(\ref{W3}) has been already defined in Eq.~(\ref{y}). Other variables are given by the following equations:
\be
\label{t}
t=\frac{1}{\omega}(p_0 E_k+\frac{m_{\phi}^2}{2}),
\ee
\vspace{.2cm}
\be 
\label{x1}
x_1=\frac{(E_k+\omega+E_m)(E_k+\omega-E_p)}{(E_k+\omega+E_p)(E_k+\omega-E_m)},
\ee
\be 
\label{x2}
x_2=\frac{(E_m-E_k+\omega)((E_k-\omega+E_p)}{(E_p-E_k+\omega)
(E_m+E_k-\omega)},
\ee
\be 
\label{x3}
x_3=\frac{(E_k-p_0+E_m)(E_k-p_0-E_p)}{(E_k-p_0-E_m)((E_k-p_0+E_p)},
\ee
\be 
\label{x4}
x_4=\frac{(E_k+p_0+E_p)(E_k+p_0-E_m)}{(E_k+p_0-E_p)(E_k+p_0+E_m)},
\ee
\be 
\label{Epm}
E_p=\sqrt{E_k^2+2 k \omega +\omega^2},~~~~~~~E_m=\sqrt{E_k^2-2 k \omega +\omega^2},
\ee
\be 
\label{Eprimbis}
E'=\frac{p_0}{2}-\frac{\omega}{2}v_{\phi},~~~~E''=\frac{p_0}{2}+\frac{\omega}{2}v_{\phi},~~~v_{\phi}=\sqrt{1-\frac{4 m_{\rm K}^2}{m_{\phi}^2}}.
\ee
Let us remark here that the terms $W_2(k)$, $W_3(k)$, $W_4(k)$, and $W_5(k)$ are singular at $\omega =0$. However, if all the terms in Eq.~(\ref{ReIr}) are added together then, due to cancellations,
the terms proportional to $1/\omega^3$, $1/\omega^2$ and $1/\omega$ vanish and the final result at $\omega=0$ is finite.


\begin{thebibliography}{99}

\bibitem{pdg2017}
  M. Tanabashi~\textit{et al.} (Particle Data Group),
   Phys. Rev.  {\bf D 98}, 030001 (2018); see Note on Scalar Mesons below 2 GeV and references given therein.

\bibitem{Klempt}
E. Klempt, A. Zaitsev, Phys. Rep. \textbf{454}, 1 (2007).  
\bibitem{Morgan}
D.~Morgan,~M.~R.~Pennington,~Phys. Rev.~\textbf{D 48}, 1185 (1993).

\bibitem{Jaffe}
R.~L.~Jaffe,~Phys. Rev.~\textbf{D 15}, 267 (1977).

\bibitem{Beveren}
E.~Van Beveren~\textit{et al.}, Z. Phys.~\textbf{C 30}, 615 (1986).

\bibitem{Johnson}
R.~L.~Jaffe,~K.~Johnson,~Phys. Lett. \textbf{B 60}, 201 (1976).

\bibitem{Weinstein}
J.~D.~Weinstein,~N.~Isgur,~Phys. Rev.~\textbf{D 41}, 2236 (1990).

\bibitem{cohen}
D. Cohen~{\it et al.}, Phys. Rev. \textbf{D 22}, 2595 (1980).
\bibitem{etkin}
A. Etkin {\it et al.}, Phys. Rev. \textbf{D 25}, 1786 (1982).

\bibitem{Ye}
  Q. J. Ye~\textit{et al.},
  Phys.\ Rev. \textbf{C 85}, 035211 (2012).
 
\bibitem{Silarski:2013rfa}
  M.~Silarski and P.~Moskal,
  Phys.\ Rev. \textbf{C 88}, 025205 (2013).

\bibitem{dzyuba}
A. Dzyuba~\textit{et al.}, Phys. Lett. \textbf{B 668}, 315 (2008).

\bibitem{bes}
M. Ablikim \textit{et al.} (BES Collaboration), Phys. Lett. \textbf{B 607}, 243 (2005).
\bibitem{Aaij:2011fx}
  R.~Aaij {\it et al.} (LHCb Collaboration),
  Phys.\ Lett. \textbf{B 698}, 115 (2011).
  
\bibitem{Nussi}
S. Nussinov and T. N. Truong, Phys. Rev. Lett. \textbf{63}, 2002 (1989).

\bibitem{Ivan}
N. N. Achasov and V. N. Ivanchenko, Nucl. Phys. \textbf{B 315}, 463 (1989).

\bibitem{Lucio1}
J. L. Lucio M. and J. Pestieau, Phys. Rev. \textbf{D 42}, 3253 (1990).

\bibitem{Bramon}
A. Bramon, A. Grau, and G. Pancheri, Phys. Lett. \textbf{B 283}, 416 (1992).

\bibitem{Bramon1}
A. Bramon, A. Grau, and G. Pancheri, Phys. Lett. \textbf{B 289}, 97 (1992).
\bibitem{Close}
F.E. Close, N. Isgur and S. Kumano, Nucl. Phys. \textbf{B 389}, 513 (1993).
  
\bibitem{Lucio2}
J. L. Lucio M. and M. Napsuciale, Nucl. Phys. \textbf{B  440}, 237 (1995).

\bibitem{KL}
N. N. Achasov, V. V. Gubin, and V. I. Shevchenko, Phys. Rev. \textbf{D 56}, 203 (1997).

\bibitem{Oller1}
J. A. Oller, Phys. Lett. \textbf{B 426}, 7 (1998).

\bibitem{Marco}
E. Marco, S. Hirenzaki, E. Oset, and H. Toki, 
Phys. Lett. \textbf{B 470}, 20 (1999).

\bibitem{Bramon2}
A. Bramon, R. Escribano, J. L. Lucio M., M. Napsuciale, and G. Pancheri,
Phys. Lett. \textbf{B 494}, 221 (2000).

\bibitem{Markushin}
V. E. Markushin, Eur. Phys. J. \textbf{A 8}, 389 (2000).

\bibitem{Aczasow2001}
N. N. Achasov and V. V. Gubin, Phys. Rev. \textbf{D 64}, 094016 (2001).


\bibitem{Oller2}
J. A. Oller, Nucl. Phys. \textbf{A 714}, 161 (2003) 

\bibitem{Kala}
Yu. S. Kalashnikova, A. E. Kudryavtsev, A. V. Nefediev, C. Hanhart, and J. Haidenbauer, Eur. Phys. J. \textbf{A 24}, 437 (2005).


\bibitem{Escribano}
R. Escribano, Phys. Rev. \textbf{D 74}, 114020 (2006).

\bibitem{Black}
D. Black, M. Harada, and J. Schechter, Phys. Rev. \textbf{D 73}, 054017 (2006).

\bibitem{Gokalp1}
A. Gokalp, C. S. Korkmaz, and O. Yilmaz, Phys. Rev. \textbf{D 75}, 013001 (2007).

\bibitem{Gokalp2}
A. Gokalp, F. Ozturk, and O. Yilmaz, Phys. Lett. \textbf{B 671}, 369 (2009).  

\bibitem{Eidelman}
S. Eidelman, S. Ivashyn, A. Korchin, G. Pancheri, and O. Shekhovtsova,
Eur. Phys. J. \textbf{C 69}, 69 (2010).

\bibitem{kloe2}
F. Ambrosino \textit{et al.} (KLOE Collaboration), Phys. Lett. \textbf{B 679}, 10 (2009).


\bibitem{kloe2008}
F.~Bossi, E.~De Lucia, J.~Lee-Franzini, S.~Miscetti, and M.~Palutan, Riv.\ Nuovo Cim.\  \textbf{31}, 531 (2008).

\bibitem{kloe0}
F. Ambrosino \textit{et al.} (KLOE Collaboration), Eur. Phys. J. \textbf{C 49}, 473 (2007).

\bibitem{kloe1}
F. Ambrosino \textit{et al.} (KLOE Collaboration), Phys. Lett. \textbf{B 681}, 5 (2009).

\bibitem{KLOE}
F. Ambrosino et al. (KLOE Collaboration), Phys. Lett. \textbf{B 634}, 148 (2006).

\bibitem{Ak1} R. R. Akhmetshin et al. (CMD-2 Collaboration), Phys. Lett. \textbf{B 462} (1999) 371.

\bibitem{Ak2} R. R. Akhmetshin et al. (CMD-2 Collaboration), Phys. Lett. \textbf{B 462} (1999) 380.

\bibitem{Ac1} M. N. Achasov et al. (SND Collaboration), Phys. Lett. \textbf{B 479} (2000) 53. 

\bibitem{Ac2} M. N. Achasov et al. (SND Collaboration), Phys. Lett. \textbf{B 485} (2000) 349. 

\bibitem{mod1}
M. Boglione, M. R. Pennington, Eur. Phys. J. \textbf{C 30}, 503 (2003).

\bibitem{NS}
G. Isidori, L. Maiani, M. Nicoli, and S. Pacetti, JHEP \textbf{0605}, 049 (2006).

\bibitem{konf}
L. Le\'sniak and M. Silarski, 
EPJ Web of Conf. \textbf{130}, 06003 (2016).

\bibitem{mod3}
N. N. Achasov, V.V. Gubin, Phys. Rev. \textbf{D 57}, 1987 (1998).
\bibitem{kam1} 

 R. Kami{\'n}ski, L. Le{\'s}niak, J. P. Maillet, Phys. Rev. \textbf{D 50}, 3145 (1994).

\bibitem{KLL}
R. Kami{\'n}ski, L. Le{\'s}niak, and B. Loiseau, Phys. Lett. \textbf{B 413}, 130 (1997).

\bibitem{kam3}
R. Kami{\'n}ski, L. Le{\'s}niak, and B. Loiseau, Eur. Phys. J. \textbf{C 9}, 141 (1999).

\bibitem{Furman}
A. Furman, L. Le{\'s}niak,  Phys. Lett. \textbf{B 538}, 266 (2002).

\bibitem{KL95} 
R. Kami{\'n}ski, L. Le{\'s}niak, Phys. Rev. \textbf{C 51}, 2264 (1995).

\bibitem{LL}
L. Le{\'s}niak, AIP Conf. Proc. \textbf{1030}, 238 (2008). 

\bibitem{L2009}
L. Le{\'s}niak, Int. J. Mod. Phys. \textbf{A 24}, 549 (2009).

\bibitem{Czyz}
H. Czy\.z, A. Grzeli\'nska, and J. H. K{\"u}hn, Phys. Lett. \textbf{B 611}, 116 (2005).

\bibitem{LST}
L. Le\'sniak and M. Silarski,
EPJ Web Conf. \textbf{166}, 00018 (2018).

\bibitem{LL96}
L. Le\'sniak, Acta Phys. Pol. \textbf{B 27}, 1835 (1996).

\bibitem{Yamaguchi}
Y. Yamaguchi and Y. Yamaguchi, Phys. Rev. \textbf{95}, 1635 (1954).

\bibitem{Oller}
J. A. Oller, E. Oset, and J. R. Pelaez, Phys. Rev. \textbf{D 59}, 074001 (1999),
Erratum: Phys. Rev. \textbf{D 60}, 099906(E) (1999), Erratum: Phys. Rev. \textbf{D 75}, 099903(E) (2007).

\bibitem{kam2}
R. Kami{\'n}ski, L. Le{\'s}niak, and K. Rybicki, Z. Phys. \textbf{C 74}, 79 (1997).

\bibitem{FL}
A. Furman, L. Le{\'s}niak,  Nucl. Phys. \textbf{B} (Proc. Suppl.) \textbf{121}, 127 (2003).

\bibitem{Bruch}
C. Bruch, A. Khodjamirian, and J.H. K{\"u}hn, Eur. Phys. J. \textbf{C 39}, 41 (2005).


\end{thebibliography}
\end{document}